# Particle Resuspension in Turbulent Boundary Layers and the Influence of Non-Gaussian Removal Forces

F. Zhang[†,*], M. Reeks[*+] and M. Kissane[†]

[†] Institut de Radioprotection et de Sûreté Nucléaire, BP 3, 13115 St-Paul-lez-Durance, France

[*] School of Mechanical and Systems Engineering, Newcastle University, Newcastle upon Tyne, NE1 7RU, UK

**Abstract**

The work described is concerned with the way micron-size particles attached to a surface are resuspended when exposed to a turbulent flow. An improved version of the Rock'n'Roll model (Reeks & Hall, 2001) is developed where this model employs a stochastic approach to resuspension involving the rocking and rolling of a particle about surface asperities arising from the moments of the fluctuating drag forces acting on the particle close to the surface. In this work, the model is improved by using values of both the streamwise fluid velocity and acceleration close to the wall obtained from Direct Numerical Simulation (DNS) of turbulent channel flow. Using analysis and numerical calculations of the drag force on a sphere near a wall in shear flow (O'Neill (1968) and Lee & Balachandar (2010)) these values are used to obtain the joint distribution of the moments of the fluctuating drag force $f(t)$ and its derivative $\dot{f}(t)$ acting on a particle attached to a surface. In so doing the influence of highly non-Gaussian forces (associated with the sweeping and ejection events in a turbulent boundary layer) on short and long term resuspension rates is examined for a sparse monolayer coverage of particles, along with the dependence of the resuspension upon the timescale of the particle motion attached to the surface, the ratio of the rms/ mean of the removal force and the distribution of adhesive forces. Model predictions of the fraction resuspended are compared with experimental results.

## 1. Introduction

The resuspension of small particles from a surface exposed to a turbulent flow occurs in a diverse range of industrial and environmental processes from clean air technology to dust storms and the spreading of crop diseases by fungal spores (see Sehmel, 1980; Nicholson, 1988). Our particular interest has been in the role resuspension can play in the release of radioactive particles following a severe accident in a range of nuclear power plant from a light-water-cooled reactor (LWR), a helium-cooled high-temperature reactor (HTR), a thermonuclear fusion reactor (e.g., ITER) and an advanced gas cooled reactor (AGR).

Our focus here is on improvements to kinetic models for the resuspension rate constant using a more detailed description of the aerodynamic removal forces generated close to a surface in a fully developed turbulent boundary layer. In particular we will consider how those forces and their time derivatives are distributed and how this influences and controls the motion of the particles on the surface and their eventual detachment and resuspension. As such we will be interested in how these improvements affect not only how many particles are removed from a surface but the rate at which they are removed and how this varies with time both in the short term (on the timescale of the removal forces) and in the much longer term. Our interest will also be in the role played by the adhesive forces, upon their magnitude and distribution and how this depends upon the particle size, surface roughness and the particle surface deformations.

[+] Corresponding author: email mike.reeks@ncl.ac.uk

The work we describe here is one of numerous studies devoted to particle resuspension over the past 30 years (see reviews by Ziskind *et al.*(1995) and Ziskind (2006)) - studies that have been devoted to measuring and computing the individual aerodynamic removal and adhesive forces and to a consideration of the various detachment mechanisms involving these forces. The adhesion studies have shown for instance that microscale roughness on average significantly reduces the adhesion compared to that for smooth contact. Particles in this case are assumed to make contact with a surface via the surface asperities common to both particle and surface. Also of importance to resuspension is the broad log normal distribution of adhesive forces that is associated with surface roughness which is significant even for a nominally smooth polished surface. As noted before (Reeks, Reed and Hall (RRH), 1988) this significantly reduces the sensitivity of resuspension to changes in flow velocity so that a threshold for resuspension is less well defined and spread out over a range of flow velocities. As for the aerodynamic forces, there is now overwhelming evidence both theoretically and experimentally that detachment from a surface is caused by a combination of the aerodynamic forces in which the drag force rather than lift force plays a dominant role (Soltani and Ahmadi (1994, 1995), Ibrahim *et al.* (2003), Guigno and Minier (2008)). It reflects the fact that the conditions for particle detachment are determined by a balance of the moments exerted by the drag and adhesive forces rather than a balance of adhesive force versus lift force (as was assumed in the early studies of resuspension) which significantly underestimates the threshold for resuspension observed experimentally (Wang, 1990). Significant work has been done in refining the components of the moment balance for both smooth and rough surfaces to account for both the influence of surface asperities and aerodynamic drag. Accordingly, it is now widely accepted that initial particle detachment from the surface occurs by rolling (Ziskind, 2006)

There has been considerable effort in incorporating these features into stochastic models that account for the influence of the near wall turbulence and coherent structures. In the early Cleaver and Yates (1975) model and the more recent Wang (1990), Soltani and Ahmadi (1994) models the removal rate is based on the frequency of so called intermittent bursts when the aerodynamic forces / moments generated during the burst exceeded some critical value. From a computational perspective, these models have been largely superseded by so called kinetic models in which the rate of removal is intimately related to the random and continuous motion of the particle-surface deformation arising from the near wall turbulence. We refer specifically to the early semi-empirical model of Wen & Kasper (1989) and the more complete model of Reeks, Reed and Hall (RRH) (1988) and the recent variants of this model due to Reeks and Hall (2001) and Vainhstein *et al.* (1997) which deal with removal initiated by rolling arising from fluctuating aerodynamics moments rather than forces. In particular the Reeks & Hall (2001) model known as the Rock'n'Roll (RnR) model considered detachment from a surface due to the rocking and rolling of a particles about surface asperities in contrast to the model of Vainhstein *et al.* (1997) which considered this about the contact area with a single asperity. Both are features of rolling on a rough surface (Ziskind et al, 1997).

We recall that the original RRH model supposed that detachment from a surface can occur by two processes: a process of energy accumulation occurring at forcing frequencies close to the resonance /natural frequency of vibration which means that a small force significantly less than the adhesive force applied at near the resonance could accumulate enough energy to overcome the surface adhesive potential barrier and detach itself from the surface; a motion far from the natural frequency where the motion of the particle attached to the surface is based upon a balance between the aerodynamic force / moments (as it varies continuously with time) and the adhesive force/ moment (as it varies with particle-surface deformation). In reality both modes of vibration (either displacement/ rotation) are implicit in the solution of the equation of motion of a single damped non linear oscillator driven by fluctuations in the aerodynamic forces due the near wall turbulence. It turns out that the degree of resonant energy transfer in all the cases of detachment considered both in practice and experiment is negligible and it is the quasi-static modes of rocking and rolling associated with the RnR and

Vainhstein *et al.* (1997) models that have been investigated and predictions compared with measurements.

With regard to measurement and benchmarking, it is important to mention the numerous experiments that have been carried out to predict the level of mechanical resuspension of deposited particles arising in LWR severe accidents. The most recent are the STORM tests which examined the resuspension of multilayer deposited aerosol particles in a pipe by high pressure dry steam flows typical of those in a LWR loss-of-coolant accident (LOCA) (Capitao and Sugaroni, 1995). As part of the STORM programme, the resuspension data were used to develop and test a number of resuspension models of various levels of sophistication. Of these, the most useful in terms of adaptability and predictability was the RnR model. The RnR model was successfully fitted to the STORM results (despite the significantly higher particle density and flow rates of these tests relative to those used to develop the model) and those of other experiments by using the data to produce values of the surface adhesion that would be consistent with the measured resuspension, Biasi *et al.* (2001).

Despite this benchmarking exercise, there are two inherent limitations of the RnR model (and by implication the Vainhstein *et al.* model). Firstly they are both based on Gaussian statistics for the aerodynamic moments / forces, despite the fact that the sequence of near wall bursting and sweeping events in a turbulent boundary layer responsible for these forces and moments are intermittent and strongly non Gaussian in nature (see e.g. Castaing *et al.* 1990). Secondly as with other stochastic models, the RnR model is strictly only applicable to the resuspension of particles on a surface with less than a monolayer coverage i.e. for particles that are essentially isolated from one another. All of the experiments associated with the measurement of resuspension particles in LWR severe accidents are concerned with resuspension from multilayer deposits of particles. In this case the removal of particles from any given layer depends upon the rate of removal from the layer above which acts as a source of uncovering and exposure of particles to the resuspending flow.

So our first objective is to take account of the non-Gaussian nature of the aerodynamic removal forces /moments in a new improved version of the RnR model. How this has been achieved and the implications for resuspension of particles from deposits that are less than monolayer coverage is the subject of the work we describe here. Our second objective is to incorporate the improved RnR model for the rate constant for resuspension from a single asperity (the primary resuspension rate constant) into a hybrid model for particle resuspension from multilayer deposits. This work is described in a subsequent paper and is based on an application of the generic model for multilayer resuspension of Friess and Yadiyaroglu (2002).

The improvements to the RnR model are based on data for the statistical fluctuations of both the streamwise fluid velocity and acceleration close to the wall obtained from direct numerical simulation (DNS) of fully-developed turbulent channel flow, translating these data into the statistical moment (couple) of the drag force and its time derivative acting on the particle attached to the surface. The RnR model is the most natural choice of model for improvement not only because of it completeness as a model and the degree to which its predictions have been benchmarked against experiment, but also because of its flexibility and computational efficiency (which makes it very suitable for incorporation into nuclear severe-accident analysis codes). We recall that the general formula for the primary resuspension rate constant depends upon the joint statistics of the aerodynamic removal force and its time derivative, but as such those statistics are purely arbitrary and not dependent on them being Gaussian.

Thus in Section 2 we begin with a description of the original RnR model and the related Vainhstein *et al.* model referring to the sources of data upon which they are based and the general formula for the primary resuspension rate constant in terms of the joint statistics of the fluctuating aerodynamic drag force and its time derivative. In Section 3 we present these distributions and show how they are calculated from a DNS of turbulent channel flow and in particular from the fluid velocities in the viscous sub-layer close to the wall. Section 4 deals with the form of the resuspension rate constant based on best fit analytic functions for the distribution of the aerodynamic drag and time derivatives extracted from the DNS data. Sections 5 and 6 deal with a comparison of the predictions of the modified RnR model with those of the original model where differences are due to the statistics and the evaluation of the aerodynamic drag used in the original model based on O'Neill's (1968) formula and the most recent DNS computations of Lee & Balachandra (2010). In particular we compare predictions with the experimental measurements of resuspension of Reeks and Hall (2001) and Ibrahim *et al.* (2003) where measurements of the adhesion due to roughness had either been measured directly (Reeks & Hall, 2001 ) or extracted from measurements of the distribution of roughness heights (Ibrahim *et al.*, 2003). We conclude in Sections 7 and 8 with a summary and discussion of the salient features of this study and some concluding remarks.

As a concluding remark, it is to be expected that the differences in statistics are most significant for the values of the primary resuspension rate constant which would be reflected in resuspension from a nominally polished surface with a very narrow distribution of adhesive forces. As this distribution broadens to that generally observed in experiment and in practice, so this difference becomes less pronounced (changes occurring logarithmically rather than linearly). Where these differences become of crucial important, as we shall show explicitly in a subsequent paper, are in the resuspension and timescales for resuspension of particles from multilayers. What we report here for sparse monolayer coverage of particles is a necessary preliminary to that consideration.

## 2. Rock'n'Roll (RnR) Quasi-static Model

This Section mainly focuses on the details of the RnR model given our choice to use it as the starting point for improving modelling. The Vainshstein *et al.* (1997) model is also outlined since it has certain similarities with the RnR model and can be easily modified in the same way as the RnR model for the purposes of comparison.

We recall that the RnR model is a stochastic model for resuspension in which particles on a microscopically rough surface, rock continuously about their points of contact with the surface roughness asperities between particle and substrate. The rocking is driven by the action of the moments of the fluctuating aerodynamic drag force acting on the particle close to the surface. Rolling is initiated when contact with the asperities is broken (point of detachment), at which point a particle is assumed to be resuspended. The detachment rate depends upon the typical timescale of the rocking motion and the concentration of particles at the detachment point. The behaviour is similar to the motion of particles in a potential well, particles escaping from the well when they have enough potential energy within the well to escape over the surface potential barrier (at the point of detachment). The motion of particles in the well takes place either quasi-statically (at a rate determined by the time scale of the turbulent aerodynamic forces) or potentially more efficiently by transfer of energy from the local turbulence to the particle motion at the natural or resonant frequency of the particle-surface deformation within the well. The quasi-static case is the one used in the current RnR model since estimates of the resonant energy transfer were found to be small (Reeks and Hall, 2001)). The motion of particles in this case can then be approximated by a force balance (or moment balance if the couple of the system is considered) between the fluctuating aerodynamic force and adhesive force (Reeks & Hall, 2001). That the quasi-static case is widely used instead of the original RnR model in nuclear severe-accident analysis codes, i.e. SOPHAEROS (Cousin et al., 2008) and AERORESUSLOG (Guentay *et al.*, 2005) is due not only to the reduction of computer processing time (since the resonant energy transfer is neglected), but also to the similar results between quasi-static and original RnR model cases.

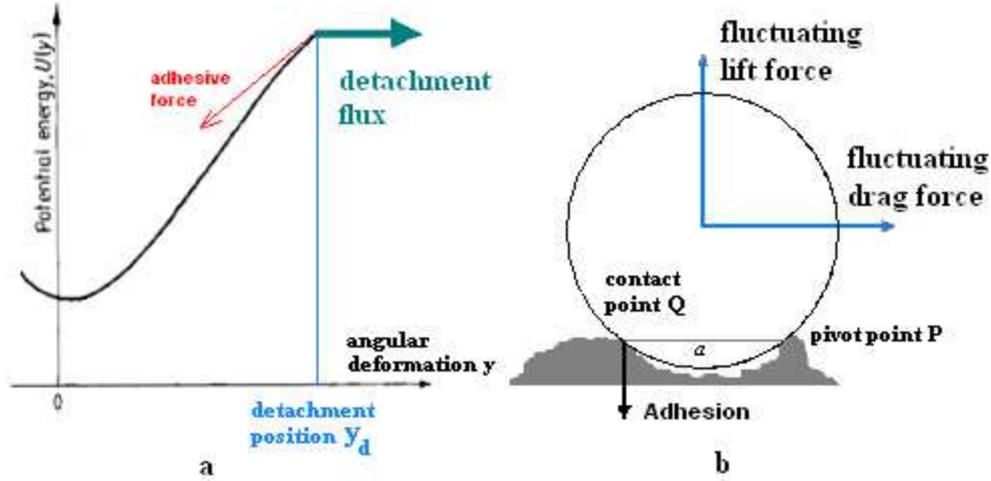

**Figure 1** - Potential well and particle couple system

The geometry of the particle-surface contact in the revised model is shown in Figure 1b in which the distribution of asperity contacts is reduced to a simple two-dimensional model of two-point asperity contact. Thus rather than the centre of the particle oscillating vertically as in the original RRH model, it will oscillate about the pivot P until contact with the other asperity at Q is broken. When this happens it is assumed that the lift force is either sufficient to break the contact at P and the particle resuspends or it rolls until the adhesion at single-point contact is sufficiently low for the particle to resuspend. In either situation the rate of resuspension is controlled by the rate at which contacts are initially broken. We note than in a recent Lagrangian stochastic model for particle resuspension (Guigno and Minier, 2008) rolling until a particle eventually resuspends has been considered in more detail.

The formula for the resuspension rate has the same form as in the original RRH model except that couples are taken account of by replacing vertical lift forces by equivalent forces based on their moments. That is, referring to Figure 1b, the equivalent force $F$ is derived from the net couple ($\Gamma$) of the system above so that

$$\Gamma = \frac{a}{2}F_L + rF_D \quad \Rightarrow \quad F = \frac{1}{2}F_L + \frac{r}{a}F_D \qquad [1]$$

where $a$ is the typical distance between asperities, $r$ the particle radius, $F_L$ the lift force and $F_D$ the drag force. The *drag amplification factor* $r/a \sim 100$ (based on Hall's experiment, Reeks & Hall 2001) meaning that drag plays the dominant role in particle removal. We recall that in the quasi-static version of the RnR model, at the detachment point (i.e. point $y_d$ in Figure 1a, referring to the angular displacement of the asperity contact at Q about P as in Fig 1b), the aerodynamic force acting on the particle (which includes the mean $<F>$ and fluctuating parts $f(t)$) balances the restoring force at each instant of time (hence the term quasi –static). So

$$\langle F \rangle + f(t) + F_A(y) = 0 \qquad [2]$$

where $F_A(y)$ is the adhesive restoring force as a function of the angular deformation $y$ of the particle. At the point of detachment ($y_d$) the adhesive pull-off force (the force required to detach the particle) is = $-F_A(y_d)$. Following the tradition of previous authors we refer to this force as the force of adhesion or the adhesive force, $f_a$. In the presence of an applied mean force $\langle F \rangle$ from Eq.[2], the value of the fluctuating component of the equivalent aerodynamic force required to detach the particle, $f_d$ is given by

$$f_d = f_a - \langle F \rangle \qquad [3]$$

So as $F(t)$ fluctuates in time, $F_A(y)$ and hence $y(t)$ change to balance it according to Eq.[2]. Every time the value of $f(t)$ exceeds the value of $f_d$ a particle is detached from the surface. So the rate of detachment depends not only on the value of $f_d$ but on the frequency with which it is exceeded, i.e., upon the typical timescale of the fluctuating aerodynamic force $f(t)$ and its distribution in time.

Based on their measurements the mean drag and lift force for a spherical particle of radius $r$ is given by Reeks and Hall (2001) as

$$\langle F_D \rangle = 32 \rho_f v_f^2 \left( \frac{r u_\tau}{v_f} \right)^2 \qquad \langle F_L \rangle = 20.9 \rho_f v_f^2 \left( \frac{r u_\tau}{v_f} \right)^{2.31} \qquad [4]$$

where $\rho_f$ is the fluid density, $v_f$ the fluid kinematic viscosity, and $u_\tau$ the wall friction velocity. The adhesive force $f_a$ is considered as a scaled reduction of the adhesive force $F_a$ for smooth contact based on the JKR model (Johnson, Kendall and Roberts, 1971). Thus

$$F_a = \frac{3}{2} \pi \gamma r \rightarrow f_a = \frac{3}{2} \pi \gamma r r'_a \qquad [5]$$

where $\gamma$ is the surface energy and $r'_a$ the normalised asperity radius $r_a / r$ where $r_a$ is the asperity radius. $r'_a$ is assumed to have a log-normal distribution $\varphi(r'_a)$ with geometric mean $\bar{r}'_a$ and geometric standard deviation $\sigma'_a$. Physically, these two parameters define the microscale roughness of the surface. $\bar{r}'_a$ is a measure of how much the adhesive force is reduced from its value for smooth contact with a surface and $\sigma'_a$ describes how broad/narrow the distribution is. For convenience we call $\bar{r}'_a$ the reduction in adhesion and $\sigma'_a$ the spread. Hall's experimental measurements of the distribution of adhesive forces on a polished stainless steel surface gave values of $\bar{r}'_a \sim 0.01$ and a spread of $\sigma'_a \sim 3$. For a log-normal distribution $\varphi(r'_a)$ is given explicitly by

$$\varphi(r'_a) = \frac{1}{\sqrt{2\pi}} \frac{1}{r'_a} \frac{1}{\ln \sigma'_a} \exp \left( - \frac{[\ln(r'_a / \bar{r}'_a)]^2}{2(\ln \sigma'_a)^2} \right) \qquad [6]$$

Biasi *et al.* (2001) took the RnR model for resuspension and added an empirical log-normal distribution of adhesive forces to reproduce the resuspension measurements of a number of experiments. Some adhesion-force parameters were tuned to fit the data of the most highly-characterised experiments, i.e., those of Hall (Reeks & Hall, 2001) and Braaten (1994). Then, for an enlarged dataset including STORM and ORNL's ART resuspension results, the best global correlations for geometric mean adhesive force and geometric spread as a function of particle geometric mean radius (in microns) were obtained, namely

$$\bar{r}'_a = 0.016 - 0.0023 r^{0.545}$$
$$\sigma'_a = 1.8 + 0.136 r^{1.4} \qquad [7]$$

where $r$ is the particle radius in microns.

The resuspension rate constant $p$, according to Reeks *et al.* (1988), is defined as the number of particles per second detached from the surface over the number of particles attached to the surface.

$$p = \int_0^\infty v P(y_d, v) dv \bigg/ \int_{-\infty}^\infty \int_{-\infty}^{y_d} P(y, v) dy dv \qquad [8]$$

where $y$ is the displacement or deformation of the centre of particle, $v = dy/dt = \dot{y}$ and $P$ is the joint distribution of $v$ and $y$. The numerator is the particle detachment flux at the point of detachment (Figure 2a) whilst the denominator is the number of particles attached to the surface, i.e. in the potential well.

Referring to Eq.[2] for the quasi-static case, we note that the angular deformation or displacement $y$ can be written as an implicit function of the fluctuating aerodynamic force, $f(t)$, i.e.

$$y(t) = \psi(f) \quad \text{and so} \quad \dot{y}(t) = \dot{f} \psi'(f) \qquad [9]$$

where $\psi'(f)$ is the first derivative of $\psi(f)$ with respect to $f$.
Then

$$p = \int_0^\infty \dot{f} P(f_d, \dot{f}) d\dot{f} \Big/ \int_{-\infty}^\infty \int_{-\infty}^{f_d} P(f, \dot{f}) df d\dot{f} \quad [10]$$

where the joint distribution $P$ of fluctuating aerodynamic force $f$ and its derivative $\dot{f}$ is assumed to be a joint normal distribution with zero correlation between the force and its derivative. Thus

$$P(f, \dot{f}) = \left[2\pi \sqrt{\langle f^2 \rangle \langle \dot{f}^2 \rangle}\right]^{-1} \exp\left(-\frac{f^2}{2\langle f^2 \rangle}\right) \exp\left(-\frac{\dot{f}^2}{2\langle \dot{f}^2 \rangle}\right) \quad [11]$$

where $\sqrt{\langle f^2 \rangle}$ is the root mean square of fluctuating force and assumed to be 0.2 of the average aerodynamic force $\langle F \rangle$.

Substituting Eq.[11] into Eq.[10], the resuspension rate constant is then given by

$$p = \frac{1}{2\pi} \sqrt{\frac{\langle \dot{f}^2 \rangle}{\langle f^2 \rangle}} \exp\left(-\frac{f_d^2}{2\langle f^2 \rangle}\right) \Big/ \frac{1}{2}\left[1 + erf\left(\frac{f_d}{\sqrt{2\langle f^2 \rangle}}\right)\right] \quad [12]$$

where $\sqrt{\frac{\langle \dot{f}^2 \rangle}{\langle f^2 \rangle}} = \omega^+\left(\frac{u_\tau^2}{\nu_f}\right) \quad [13]$

$\omega^+$ is the value of $\sqrt{\langle \dot{f}^2 \rangle / \langle f^2 \rangle}$ in wall units and represents the typical frequency of particle motion (in radians/s) in the surface adhesive potential well (Note: $\omega^{+-1}$ is not the timescale of $f(t)$). In the original RnR model $\omega^+$ is 0.0413. Note that for an harmonic oscillator $f(t) = e^{i\omega t}$, $\sqrt{\langle \dot{f}^2 \rangle / \langle f^2 \rangle}$ is identical to $\omega$.

For particles with less than a monolayer coverage on a surface, the fraction remaining $f_R(t)$ and the fractional rate of resuspension $\Lambda(t)$ at time $t$ are given by

$$f_R(t) = \int_0^\infty \exp[-(p(r_a'))t] \varphi(r_a') dr_a'$$

$$\Lambda(t) = -\dot{f}_R(t) = \int_0^\infty p(r_a') \exp[-(p(r_a'))t] \varphi(r_a') dr_a'$$

[14a]

### The Vainshstein et al.(1997) model for resuspension

We mention here the salient features of this model because it is similar to the RnR model described above and we can easily compare predictions with those of the RnR model with the same non-Gaussian statistics by simply changing the value of the drag amplification factor, $r/a$ in the formula for the effective removal force $F$ in Eq.(1). Thus instead of rocking about several asperities, a particle rocks about the contact area of a single asperity, so the effective moment arm of the adhesive moment is the contact radius.

The adhesive moment is given by

$$M_a = 12.5 \frac{\gamma^{4/3} r_a^{5/3}}{\kappa^{1/3}}$$

where $\kappa$ is the elastic constant (composite Young's modulus) defined by

$$\kappa = \frac{4}{3}\left(\frac{1-\nu_1^2}{E_1} + \frac{1-\nu_2^2}{E_2}\right)^{-1}$$

where $v_1$, $v_2$ are Poisson's ratio for the particle and the substrate, respectively, and $E_1$, $E_2$ are Young's moduli. The drag moment is given by

$$M_D = 1.399 F_D r$$

At detachment (onset of rolling) $M_D = M_a$ which means that the effective amplification of the drag force, $r/a$ in this model is given by

$$r/a = 1.399(rf_a/M_a) = 0.5274 \frac{r\kappa^{1/3}}{\gamma^{1/3} r_a^{2/3}} \qquad [14b]$$

where we have replaced $f_a = 1.5\pi\gamma r_a$. So once the roughness is defined in terms of a ratio of $r_a/r = r'_a$, then this amplification factor varies as $r'_a$ and the radius of the particle and is therefore a random quantity in the same way as the surface adhesive force is a random variable over the entire surface. In the RnR model involving the rocking about asperities, the value of $a$ is regarded as the typical distance between asperities and related to surface roughness geometry in a more general way than directly related to $r'_a$ as in Eq.(14b). Its value was measured in Hall's experiment (Reeks and Hall, 2001) as a single measure of the entire distribution of surface roughness and used legitimately in the RnR model to predict the resuspension measured in the resuspension phase of the experiment with the same particle-surface combination. Using a value of $r/a = 100$ attributed to the whole surface was not meant to imply that this was a fixed value but only typical of the polished surface used in the Hall experiment. As one might expect this ratio would be expected to vary from particle to particle on the surface in much the same way as the rms roughness varies beneath each particle.

## 3. Statistics of fluctuating aerodynamic removal forces based on a DNS of turbulent channel flow

In this Section, we present and show how the distributions of the fluctuating aerodynamic force and its time derivative (assumed to be normally distributed in the original RnR model), are calculated from a Direct Numerical Simulation of turbulent channel flow and in particular the fluid velocities in the viscous sublayer close to the wall. We show how the measurement of the streamwise velocities are converted into values for the drag force acting on a spherical particle attached to the wall and how we use this data not only to determine the distributions of $f$ and $\dot{f}$ but also revised values for the typical forcing frequency $\omega^+$ and the ratio $f_{rms}$ ($\langle f^2 \rangle^{1/2} / \langle F \rangle$). Analytic forms for the distributions are fitted to the 'measured' DNS distributions which are then compared to the Gaussian forms used in the original model.

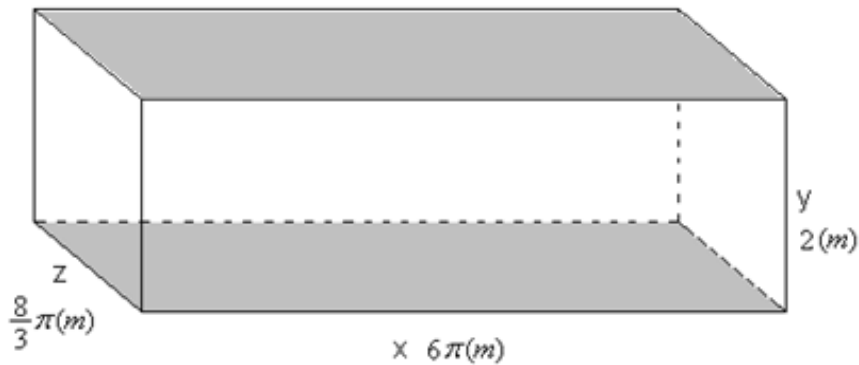

**Figure 2** - Domain of DNS calculation

A spectral projection method for incompressible flow simulation based on an orthogonal decomposition of the velocity into two solenoid fields (Buffat *et al.*, 2011) is applied for DNS. The approximation is based on Fourier expansions in the streamwise (x) and spanwise (z) directions and an orthogonal expansion of Chebyshev polynomials (proposed by Moser *et al.*, 1983) in the wall normal (y) direction in order to satisfy the wall boundary conditions. The boundary conditions are no-slip on top and bottom walls and periodic in the streamwise and spanwise directions.

|     | x  | y | z      | δ | grid          | $Re_\tau$ | $\Delta t^+$ | steps |
|-----|----|---|--------|---|---------------|-----------|--------------|-------|
| DNS | 6π | 2 | ⅔ π    | 1 | 384 x 193 x 384 | 180       | 0.0336       | 63738 |

**Table 1 -** Simulation parameters in DNS

The fluid instantaneous streamwise velocity $u$ was obtained for different values of $y^+$ away from the wall at each time step. Assuming the local fluid velocity is similar to the particle velocity, the instantaneous drag forces acting on the particle is then calculated from the velocities using O'Neill's (1968) formula which is derived from a simple drag force solution of the Stokes flow equation via Fourier–Bessel transforms for a sphere sitting on the wall in a viscous sub-layer, namely

$$F_D = 1.7 \cdot 6\pi \mu_f r u = 10.2\pi \frac{r^+ \mu_f^2 \rho_f}{u_\tau} u \qquad [15]$$

where $r$ in turn is the distance of the centre of the spherical particle from the wall and $r^+$ is the dimensionless particle radius which is considered as $y^+$ from the wall. Since in the RnR model the effective drag force through its moment makes the major contribution to the aerodynamic force (the drag force is multiplied by a factor of 100 and the lift force is reduced to half, following Eq.[1]), the lift force contribution has been neglected. The aerodynamic force contains a mean and a fluctuating component, i.e. given by

$$f = F - \langle F \rangle \qquad [16]$$

Here $F$ and $f$ are aerodynamic forces after being scaled up (Eq.[1]). The time derivative of the fluctuating aerodynamic force $\dot{f}$ is calculated by the first-order method,

$$\dot{f}_i = \frac{f_{i+1} - f_i}{\Delta t} \qquad [17]$$

Let $z_1$ and $z_2$ be the fluctuating aerodynamic force and derivative normalized on their r.m.s. values, so

$$z_1 = \frac{f}{\sqrt{\langle f^2 \rangle}}, \quad z_2 = \frac{\dot{f}}{\sqrt{\langle \dot{f}^2 \rangle}} \qquad [18]$$

The histograms of $z_1$ and $z_2$ are shown below for the case $y^+ = 6$, indicating that for the DNS data the distribution of $z_1$ fits a Rayleigh distribution (Figure 3.I) and that for $z_2$ a Johnson SU distribution (Figure 3.II).

We should clarify that, while the Rayleigh fit is imperfect around the small-fluctuations point (Fig. 3.I), it captures very well the distribution of larger-amplitude fluctuations. It is therefore very much closer to reality (i.e., to the DNS data) than the original Gaussian assumption. This is discussed again below

As shown in Figure 3.I, compared to the Gaussian case which is assumed in the original RnR model, the graph has a positive skewness (for $y^+ = 6$, skewness = 0.568), in other words, there is a significant contribution in the tails of the Rayleigh distribution compared to a Gaussian.

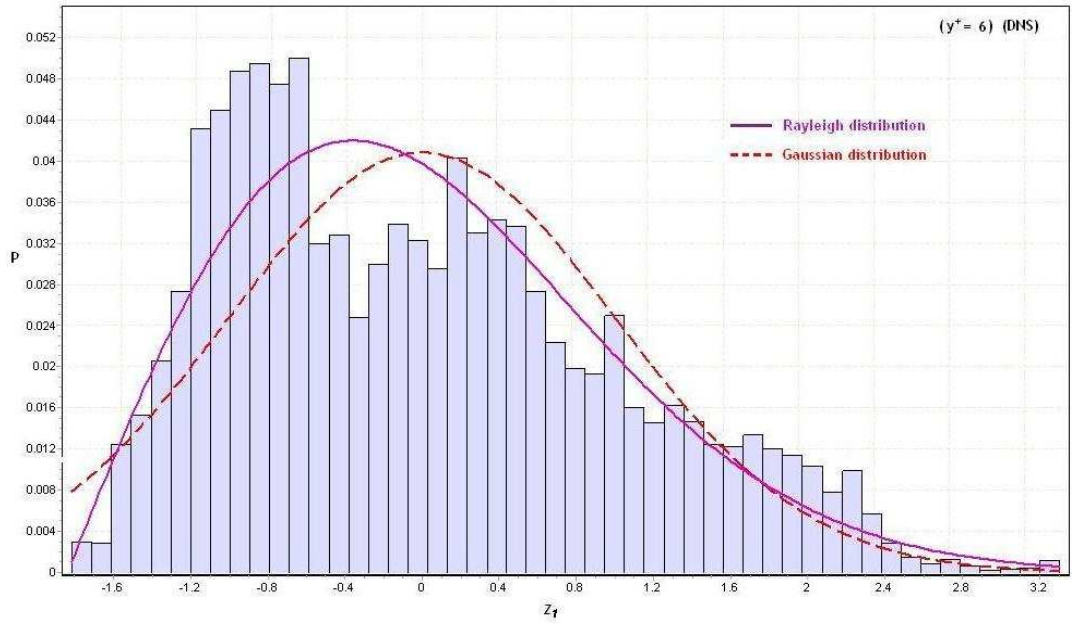

I

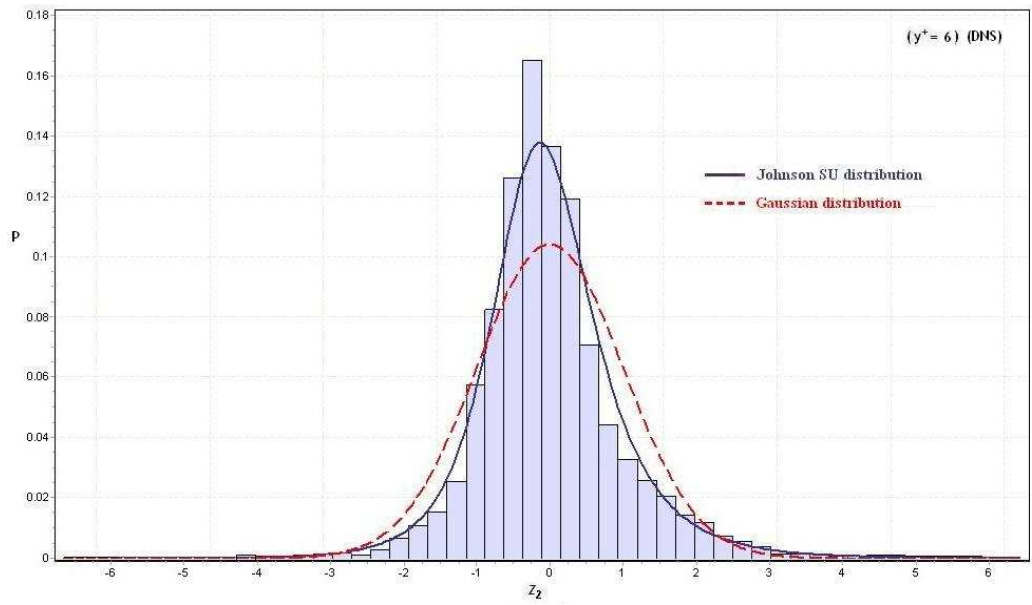

II

**Figure 3 -** Distribution of normalized fluctuating aerodynamic force (top, I) and its derivative (bottom, II) (DNS statistics at $y^+ = 6$)

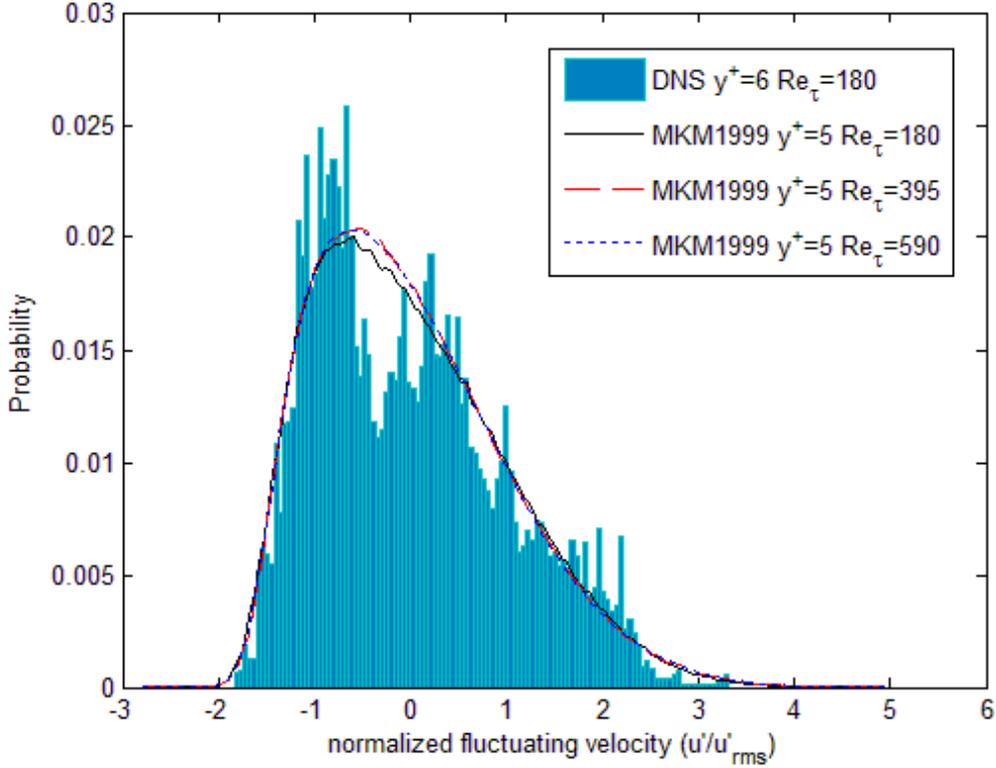

**Figure 4** - Histogram of normalized streamwise fluctuating velocity obtained from our DNS data compared to that from DNS of Moser *et al.* (1999)

Surprisingly there are only a few measurements of the distribution of the fluctuating fluid velocities in the near wall region reported in the literature. Figure 4 shows the histogram obtained from our DNS calculations compared to that obtained from the DNS data of Moser *et al.* (1999) for different wall Reynolds numbers. Similarly to above with respect to the Rayleigh representation, we note that the dip in the region of small-amplitude fluctuations in our DNS histogram is absent from the Moser *et al.* data but that the histograms are very close to one another in the wings (i.e., for large-amplitude fluctuations). The dip in our data may be a real effect or due to a lack of velocity data in that region (near 0). However, in an exercise not presented here, our results of comparison of the modified model with Rayleigh distribution and modified model with raw data distribution show that these very small fluctuations have negligible effect on resuspension. Therefore, the Rayleigh distribution is used to fit the histogram of fluctuating aerodynamic force.

## 4. Modification of resuspension rate constant

In the original model the fluctuating aerodynamic force $f$ and its derivative $\dot{f}$ are assumed to be statistically independent of each other with a normal distribution. The statistical independence is based on the fact that in steady state $\langle f \dot{f} \rangle = \tfrac{1}{2} \dfrac{d}{dt} \langle f^2 \rangle = 0$. We make the same assumption for the non-Gaussian forces, i.e., that $z_1$ and $z_2$ are statistically independent of one another with a joint distribution compounded of a Rayleigh distribution for $z_1$ and a Johnson SU distribution for $z_2$. More precisely

$$P(z_1, z_2) = \frac{z_1 + A_1}{A_2^2} \exp\left(-\frac{1}{2}\left(\frac{z_1 + A_1}{A_2}\right)^2\right) \cdot \frac{B_1}{B_2 \sqrt{2\pi} \sqrt{z^2+1}} \exp\left(-\frac{1}{2}\left(B_3 + B_1 \ln\left(z + \sqrt{z^2+1}\right)\right)^2\right)$$

[19]

where $A_1$, $A_2$, $B_1$, $B_2$, $B_3$ and $B_4$ are constants depending on the wall distance $y^+$ and

$$z = \frac{z_2 - B_4}{B_2}$$

Substitute Eq.[19] into Eq.[10], the modified resuspension rate constant is obtained

$$p = \sqrt{\frac{\langle \dot{f}^2 \rangle}{\langle f^2 \rangle}} \int_0^\infty z_2 \, P(z_d, z_2) dz_2 \Big/ \int_{-\infty}^\infty \int_{-\infty}^{z_d} P(z_1, z_2) dz_1 dz_2$$

$$= B_{\dot{f}} \sqrt{\frac{\langle \dot{f}^2 \rangle}{\langle f^2 \rangle}} \frac{z_d + A_1}{A_2^2} \exp\left(-\frac{1}{2}\left(\frac{z_d + A_1}{A_2}\right)^2\right) \Big/ \left[1 - \exp\left(-\frac{1}{2}\left(\frac{z_d + A_1}{A_2}\right)^2\right)\right] \quad [20]$$

where $z_d = \frac{f_d}{\sqrt{\langle f^2 \rangle}}$.

In the original RnR model the term $\sqrt{\langle f^2 \rangle}$ is expressed as a fraction $f_{rms}$ of the mean aerodynamic force, i.e.,

$$\sqrt{\langle f^2 \rangle} = f_{rms} \langle F \rangle \quad [21]$$

Note that the value in the original RnR model $f_{rms}$ was taken as 0.2 based on Hall's measurements (Reeks et al., 1988).

As in Eq.[13] for the original RnR model we write

$$\sqrt{\frac{\langle \dot{f}^2 \rangle}{\langle f^2 \rangle}} = \omega^+ \left(\frac{u_\tau^2}{\nu_f}\right) \quad [22]$$

Values for the various dimensionless parameters associated with the formula for $p$ in Eq.[20] are given in Table 1 for values of $y^+ = 0.1, 0.6, 2$ and 6 from the DNS data.

|  | $B_{\dot{f}}$ | $A_1$ | $A_2$ | $\omega^+$ | $f_{rms} = \sqrt{\langle f^2 \rangle}/\langle F \rangle$ |
|---|---|---|---|---|---|
| $y^+ = 0.1$ | 0.3437 | 1.8126 | 1.4638 | 0.1642 | 0.366 |
| $y^+ = 0.6$ | 0.3469 | 1.7848 | 1.4466 | 0.1520 | 0.366 |
| $y^+ = 1.9$ | 0.3512 | 1.7599 | 1.4313 | 0.1313 | 0.365 |
| $y^+ = 6$ | 0.3586 | 1.8361 | 1.4784 | 0.1271 | 0.346 |

**Table 2** - Values of parameters used in the formula for resuspension rate constant $p$

From the table above one can observe from DNS results that in the viscous sublayer ($y^+ < 6$), the statistics of the fluctuating resultant force and its time derivative (normalized on their rms values) are almost independent of $y^+$ (parameters $B_{\dot{f}}$, $A_1$ and $A_2$ are very close). Also the value of the rms coefficient $f_{rms}$ is almost independent of $y^+$ and with approximately the same values for the DNS measurements. By contrast $\omega^+$ increases when $y^+$ decreases. In the modified model, the parameters for $y^+ = 0.1$ are used since it is the region closest to the wall. The importance of the parameters $B_{\dot{f}}$, $A_1$ and $A_2$ that define the non-Gaussian distributions as distinct from a Gaussian distribution and the two parameters ($f_{rms}$ and $\omega^+$) on resuspension will be investigated in the subsequent analysis and figures given below.

## 5. Analysis of Results

In this Section, we compare the predictions of the modified RnR model based on DNS data with those of the original RnR model. The difference depends on 3 distinguishable contributions: the distributions of $f$ and $\dot{f}$ (normalised on their rms values) and the different values of $\omega^+$ and $f_{rms}$. First the difference between the Gaussian and non-Gaussian models is examined in terms of the influence the Gaussian and non-Gaussian distributions for the removal forces have on the resuspension rate which reflect the relative contribution from the highly non-Gaussian forces (associated with the sweeping and ejection events in a turbulent boundary layer). It will show how the dependence of the resuspension rate constants on the adhesive force is dependent upon a Gaussian distribution in the Gaussian model and upon a Rayleigh distribution in the non-Gaussian model. We then compare predictions of the original Gaussian RnR model with those of the modified non-Gaussian RnR model based on the DNS results for $y^+ = 0.1$ in Table 2 where the difference also depends upon the different values of $\omega^+$ and the values of $f_{rms}$ (the ratio of the rms of the aerodynamic removal force to its mean value). In this case we shall compare predictions with the experimental results in the Hall experiment (Reeks and Hall, 2001) and with those reported in Ibrahim et al (2003).

As a preliminary to these comparisons it will be found useful to introduce a few relevant scaling parameters and relationships. They will allow us to make the comparisons more universal and independent of particular flow situations. We recall that the normalized fluctuating resultant force at the detachment point ($z_d$), is defined as

$$z_d = \frac{f_a - \langle F \rangle}{\sqrt{\langle f^2 \rangle}} = \frac{F_a r_a' - \langle F \rangle}{f_{rms} \langle F \rangle} \qquad [23]$$

where $F_a$ is the adhesive force for smooth contact Then the normalized adhesive force (or the ratio of adhesive force to the rms of the aerodynamic force) is given by

$$z_a = z_d + \frac{1}{f_{rms}} = \frac{F_a r_a'}{f_{rms} \langle F \rangle} \qquad [24]$$

where $r_a'$ is the normalized asperity radius which is assumed to have a log-normal distribution $\varphi(\overline{r_a'}, \sigma_a')$. This means that $z_a$ is also distributed as a log-normal distribution. The geometric mean is defined as

$$\overline{z}_a = \frac{F_a \overline{r_a'}}{f_{rms} \langle F \rangle} = \frac{{}^3\!/\!_2 \pi \gamma r}{f_{rms} \langle F \rangle} \overline{r_a'} \qquad [25]$$

We note that $z_a / \overline{z}_a = r_a' / \overline{r_a'}$ so the geometric spread $\sigma_a'$ for $r_a'$ is the same as that for $z_a$.

Thus the log-normal distribution $\varphi(\overline{r_a'}, \sigma_a')$ can be replaced by $\varphi(\overline{z}_a, \sigma_a')$ for the distribution of adhesive forces.

The resuspension rate constant $p$ is a function of $z_d$. Thus the particle fraction remaining on the surface and the fractional resuspension rate at time $t$ are given respectively by

$$f_R(t) = \int_0^\infty \exp\left[-p(z_d)t\right] \varphi(z_a) dz_a$$
$$\Lambda(t) = -\dot{f}_R(t) = \int_0^\infty p(z_d) \exp\left[-p(z_d)t\right] \varphi(z_a) dz_a \qquad [26]$$

It is noted that $\omega$ is the typical forcing frequency of the particle in the potential well, defined as

$$\omega = \sqrt{\frac{\langle \dot{f}^2 \rangle}{\langle f^2 \rangle}} = \omega^+ \left(\frac{u_\tau^2}{\nu_f}\right) \qquad [27]$$

So that $\omega^{-1}$ is a natural time scale for the resuspension, and the resuspension rate ($\Lambda$), resuspension rate constant ($p$) and the resuspension time ($t$) can be usefully normalized on this typical frequency $\omega$. Thus

$$\hat{\Lambda} = \Lambda/\omega, \quad \hat{p} = p/\omega, \quad \hat{t} = \omega t \qquad [28]$$

The normalized resuspension rate is then given as

$$\hat{\Lambda}(t) = \int_0^\infty \hat{p}(z_d) \exp\left[-\hat{p}(z_d)\hat{t}\right] \varphi(z_a) dz_a$$

This means that for a given value of $\omega t$, resuspension rates scale on $\omega$ (and hence $\omega^+$) since $\hat{\Lambda}$ will be independent of $\omega$. We note also that the fraction resuspended will be the same at times $t$ for which $\omega t$ has a constant value. For any given value of $\omega$ the fraction resuspended will increase with increasing $\omega$ until a point of saturation is reached in the limit $\omega t = 1$ when the dependence on $\omega$ is reduced to zero. See Zhang (2011) for confirmation and further details.

## 5.1 Gaussian vs. Non-Gaussian Distribution (DNS)

In this Section we will compare the predictions using a non-Gaussian model for the resuspension rate constant $p_{nG}$ based on Eq.[20] with those obtained using a Gaussian model. The values of the constants in Eq.[20] are those given in Table 2. For the Gaussian model the resuspension rate constant $p_G$ is given by

$$p_G(z_d) = \tfrac{1}{2\pi}\omega \exp\left(-\tfrac{1}{2}z_d^2\right)$$

where $\omega = \sqrt{\langle \dot{f}^2 \rangle / \langle f^2 \rangle}$ and $z_d = (f_a - \langle F \rangle)/\sqrt{\langle f^2 \rangle}$. For future reference we shall also use the normalised adhesive force $z_a = f_a / \sqrt{\langle f^2 \rangle}$ so that $z_d = z_a - f_{rms}^{-1}$ because, unlike $z_d$, a log-normal distribution of asperity radii corresponds to a log-normal distribution of $z_a$ with the same geometric spread (see Eq.[24] and [25]).

In comparing the non-Gaussian and Gaussian models we shall naturally use the same value of $\omega = \sqrt{\langle \dot{f}^2 \rangle / \langle f^2 \rangle}$ and $f_{rms}$. In fact we shall plot the results so that the differences are independent of the value of $\omega$ reflecting only the difference between a Gaussian and non-Gaussian distribution of fluctuating aerodynamic forces with the same standard deviation (to be more precise, a Gaussian with a Rayleigh distribution). Later on we will compare the predictions based on the original RnR model with those based on the non-Gaussian resuspension rate (which we refer to as the modified RnR model), but in these cases the values of $\omega$ and $f_{rms}$ are different.

To begin with we compare the values for the resuspension rate constant for the Gaussian and non-Gaussian models when the adhesive force balances the mean aerodynamic force, i.e., $f_a$ or $z_d = 0$. For the Gaussian model this value corresponds to the maximum value of the resuspension rate constant. For a Gaussian model (as in the original RnR model),

$$p_G(0) = \tfrac{1}{2\pi}\omega = 0.15915\omega$$

This is also the maximum value and applies for $z_d < 0.75$. We recall that using Hall's measurements for the original RnR model $p_G(0) = 0.00658 u_\tau^2 / \nu_f$.

In the case of the non–Gaussian model,

$$p_{nG}(0) = B_{\dot{f}}\left(\frac{A_1}{A_2^2}\right)\exp\left(-\frac{1}{2}\left(\frac{A_1}{A_2}\right)^2\right)\left(1 - \exp\left(-\frac{1}{2}\left(\frac{A_1}{A_2}\right)^2\right)\right)^{-1}\omega$$

which, using the values for $A_1$, $A_2$, $B_{\dot{f}}$ of $y^+ = 0.1$ given in Table 2, gives

$$p_{nG}(0) = 0.25223\omega$$

We note from Figure 5 that $p_{nG}(z_d) > p_G(z_d)$ for $z_d < 0.5$ because the maximum value of the resuspension rate constant in the Gaussian model is set at $p_G(0)$ as in the original model. Note the negative skewness of the distribution of aerodynamic forces means there are more particles on the surface which experience forces < the mean removal force $\langle F \rangle$ than those > $\langle F \rangle$. However as shown in Figure 5 as $z_d$ increases beyond 0.5, the difference between Gaussian and non-Gaussian decreases until at $z_d \approx 2.1$ they are both the same. Beyond this value, the non-Gaussian rate constant exceeds the Gaussian value. Particularly striking is the large difference between the two predictions for values of the resuspension rate constant for

$z_d$ ? 1 which although $= p_{nG}(0), p_G(0)$, reflects the significant difference between the two distributions for aerodynamic removal forces in the wings of the distribution (corresponding to the highly intermittent bursting and sweeping events of fluid motion near the wall).

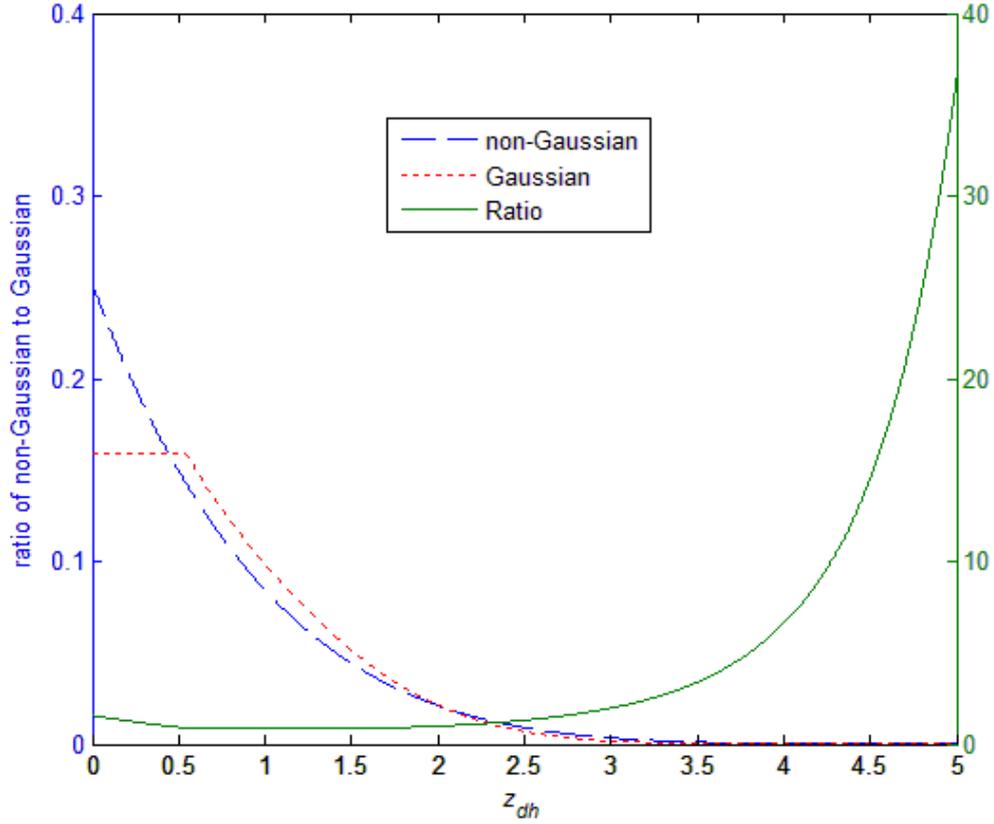

**Figure 5** - Normalized resuspension rate constant between non-Gaussian and Gaussian

The DNS measurements are only reliable out to $z_d \approx 4$, but even so from Figure 5, the ratio of $p_{nG} / p_G \approx 6$. The form of the distribution for values of $z_d > 4$ would seem to indicate the difference between the two predictions increases significantly.

It is interesting to see how this significant difference in the values of the rate constants for the two models for large values of the adhesive force is reduced when in practice we generally have a broad spread of adhesive forces. To show this, we effectively plot the ratio of the initial resuspension rate as a function of geometric mean of the normalized adhesive force, $z_a$ for various values of the spread (Figure 6) and then the same ratio as a function of the spread for a large value of the geometric mean (Figure 7). Note that a log-normal distribution of normalized asperity radii will have the same spread as a log normal distribution of normalized adhesive forces (as shown before). For a very narrow spread ~ 1.01 we would expect to reproduce the ratio of resuspension rate constants shown in Figure 5. However as the spread increases so the relative importance and contribution to the resuspension rates from the higher values of the normalized adhesive force $z_a$ is markedly less, even when the geometric mean of $z_a$ ~ 8 (note that for comparison with Figure 5 for a value of $z_a$ = 8, $z_d$ ~ 5.27 for a value of $1/f_{rms}$ ~ 2.73 based on the value of $f_{rms}$ = 0.366 in Table 2). In fact for a spread of 2 (nominally smooth surfaces), the ratio is less for large values of the geometric mean of the normalized adhesive force compared to its value for zero geometric mean of $z_a$.

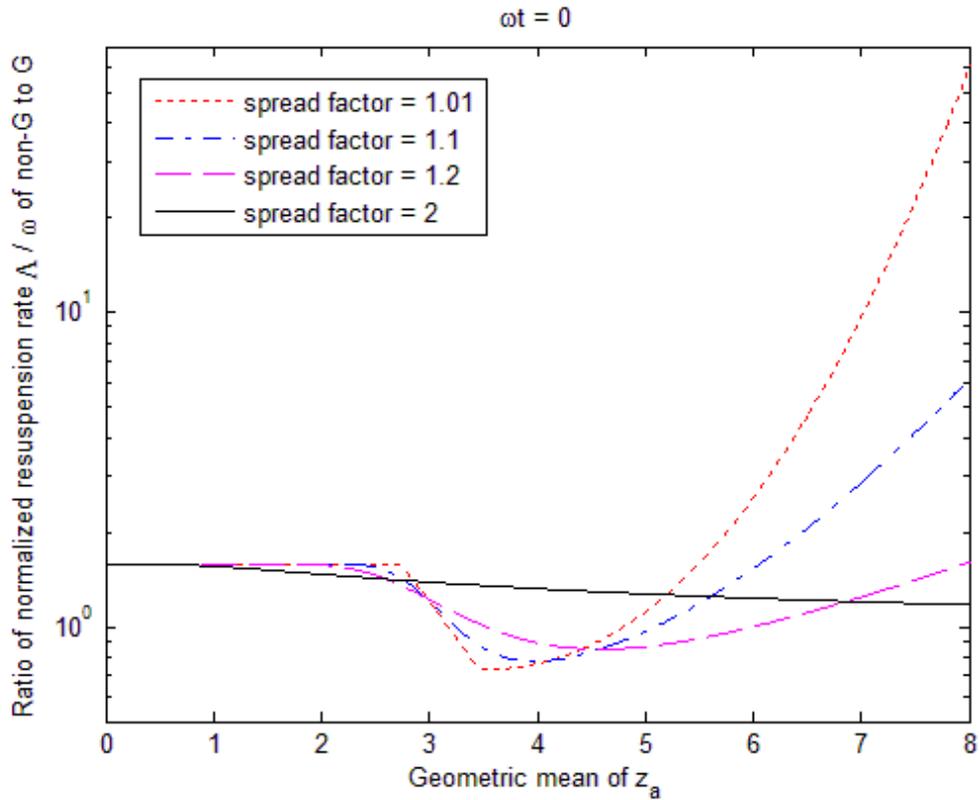

**Figure 6** - Ratio of normalized initial resuspension rate of non-Gaussian to Gaussian vs. Geometric mean of $z_a$ (ratio of adhesive force $f_a$ / rms of fluctuating aerodynamics force $\sqrt{\langle f^2 \rangle}$ )

Figure 7 shows the sensitivity of the ratio of normalized resuspension rates to changes in the spread for a large value of the geometric mean of $z_a = 8$. Note the ratio drops to unity for a spread as narrow as 1.2 and actually drops below unity but flattens out to a value ~ 1.5 as the spread increases. All this reflects the regions where the ratio of the rate constants is less than 1 for values of $z_a$ between 2.73 (when mean aerodynamic forces $\approx$ adhesive force) and 5 and $z_a > 5$ when the ratio > 1 and the relative contributions these regions of the curve of the resuspension rate constant make to the overall net resuspension rate. Of course resuspension is not an instantaneous process and we know that the resuspension rates will vary significantly in the short term for $0 < \omega t < 10$ to $\omega t$ ? 1 in the long-term.

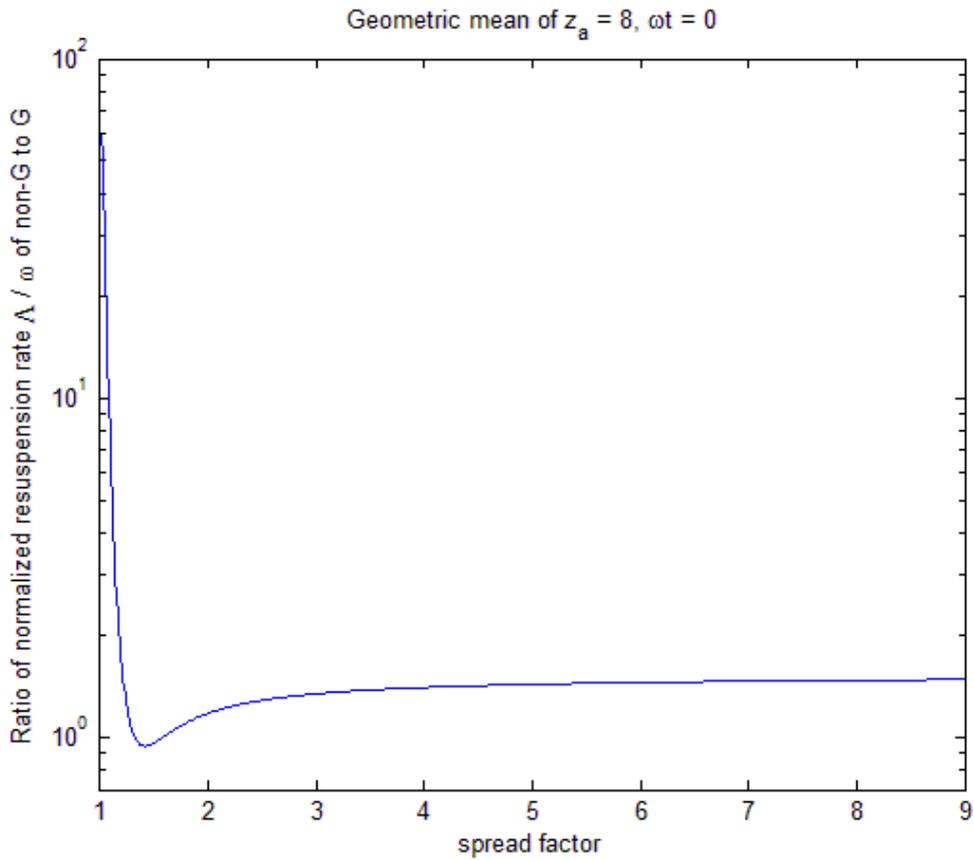

**Figure 7** - Ratio of normalized resuspension rate of non-Gaussian to Gaussian vs. spread

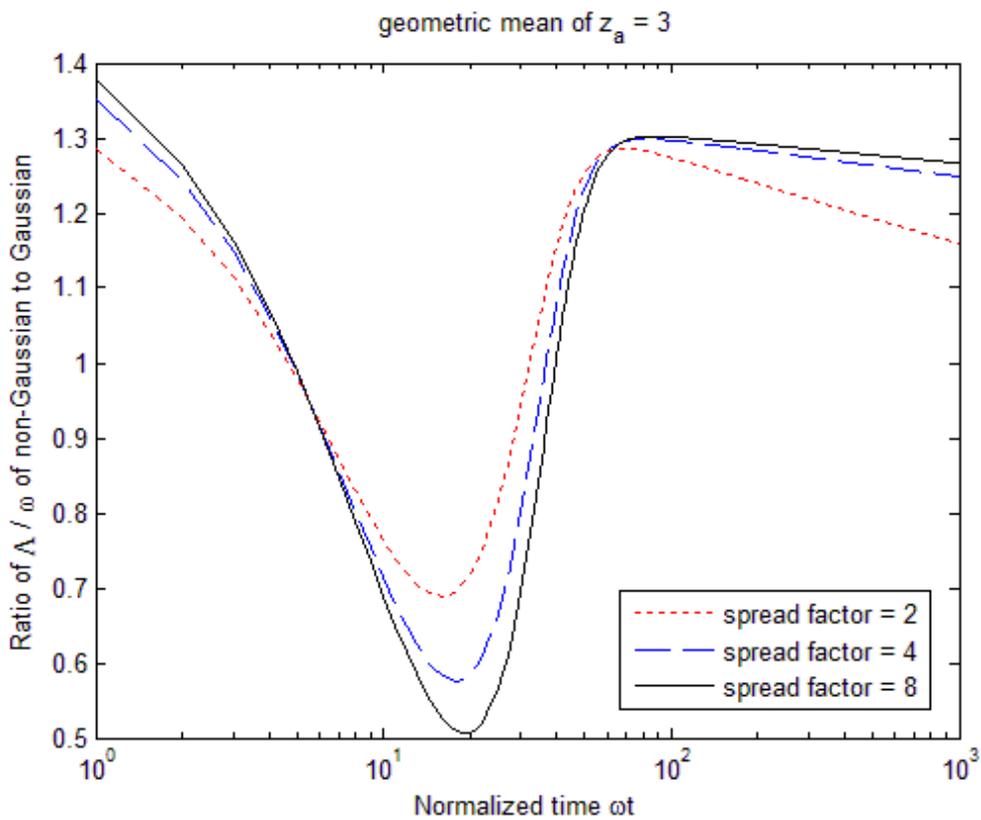

**Figure 8** - Ratio of normalized resuspension rate of non-Gaussian to Gaussian vs. $\omega t$

Figure 8 shows the ratio of non-Gaussian to Gaussian normalized resuspension rates as a function of time for a value of the geometric mean of the normalised adhesive forces $z_a = 3 \approx$ mean aerodynamic force ($z_a \sim 3$) which turns out to be typical for the range of values of the geometric mean from 3 to 8. Thus for an adhesive spread from 2 to 8, the ratio starts off $> 1$ (as in Figure 6) decreases to a value close to unity at $\omega t \sim 5$, and reaches a minimum value for value of $\omega t \sim 20$ (the precise value increasing with the adhesive spread). The actual minimum value is less the greater the spread. In the region of $5 < \omega t < 40$, the ratio is less than 1 and for $\omega t > 40$ the ratio is greater than 1 and rising to a maximum value $\sim 1.3$ at $\omega t \sim 80$. Beyond this value of $\omega t$, the ratio flattens out to a constant value larger than 1 which depends on the spread factors and geometric means of $z_a$. It shows that for the long-term, the resuspension rate of the non-Gaussian model is always larger than the Gaussian case at a fix ratio value. In Figure 9 we show the actual values of the resuspension rates for the Gaussian and non-Gaussian models indicating the transition from short to long-term resuspension occurring at $\omega t > 50$.

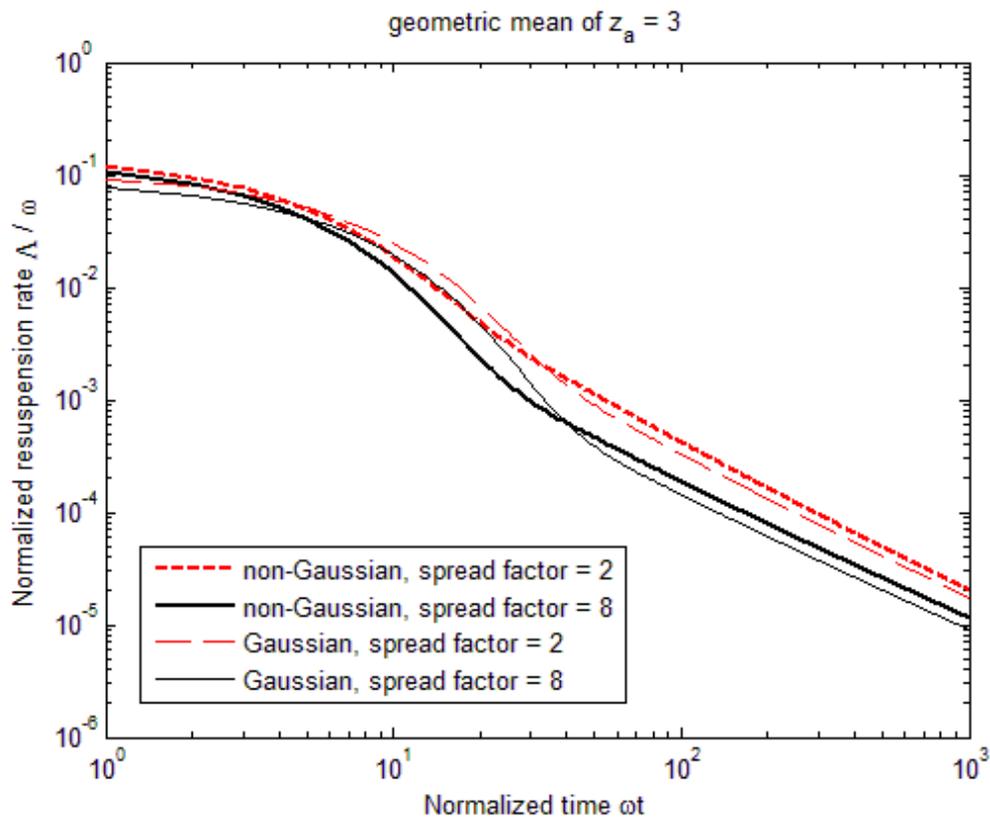

**Figure 9** - Normalized resuspension rate of non-Gaussian and Gaussian model vs. $\omega t$

## 5.2 Comparison of the Long Term Resuspension Rates

We shall now consider how the two parameters $\omega^+$ and $f_{rms}$ affect the long term resuspension rate. Starting with $f_{rms}$, Figure 10 shows the comparison of normalized resuspension rate (normalized on $\omega$ so the parameter $\omega^+$ is not considered) of the original RnR and modified models. So in these two models, apart from the difference of joint Gaussian versus non-Gaussian distributions for $f(t)$ and $\dot{f}(t)$, $f_{rms}$ is the only influence on the normalised resuspension rate $\Lambda/\omega$ (original: $f_{rms} = 0.2$; modified: $f_{rms} = 0.366$). Compared to Figure 9, the difference between the long term resuspension rate of modified and original model in Figure 10 is much greater for the same spread factor. Therefore, the value of the parameter $f_{rms}$ (the ratio of rms of the force to its mean value) has a significant influence on the value of the long term resuspension rate.

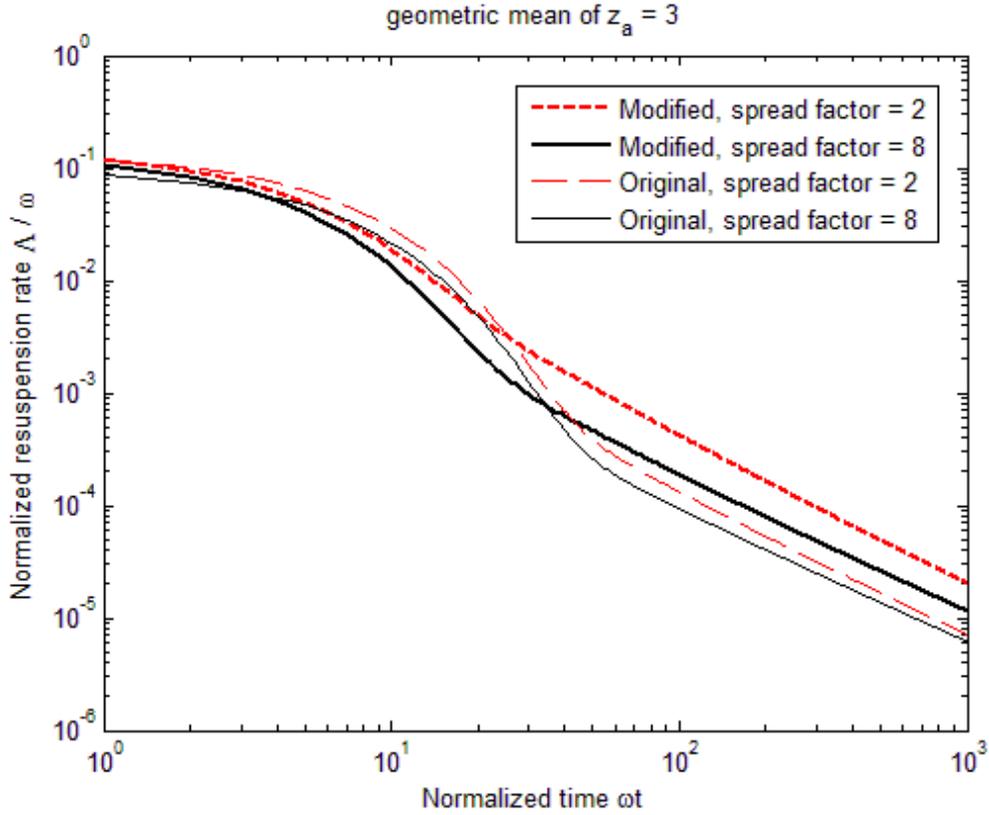

**Figure 10** - Normalized resuspension rate of modified and original model vs. $\omega t$

We can clearly use the form for the normalised resuspension rates in Figure 10 to obtain the dependence of $\Lambda(t)$ on $\omega$, namely

$$\Lambda(t) = \omega\, \hat{\Lambda}(\omega t)$$

In the short term ($\omega t = 1$), therefore, the resuspension rate scales directly as $\omega$. However in the long term the influence of $\omega$ is significantly reduced. This can be illustrated best by recalling that that the long-term resuspension rate follows a power law decay of the form (Reeks *et al.*, 1988),

$$\Lambda(t) = \xi_1 t^{-\xi_2} \qquad [30]$$

where $\xi_1$ and $\xi_2$ are constants with $\xi_2 \approx 1$ but $\neq 1$.

This implies that the corresponding normalized resuspension rate $\hat{\Lambda}$ behaves as

$$\hat{\Lambda}(\hat{t}) = \hat{\xi}_1 \hat{t}^{-\hat{\xi}_2} \ \text{ with } \ \hat{\Lambda} = \Lambda/\omega \ \text{ and } \ \hat{t} = \omega t \qquad [31]$$

Combining Eq.[30] and Eq.[31], we have

$$\xi_1 = \hat{\xi}_1 \omega^{1-\hat{\xi}_2} \quad \xi_2 = \hat{\xi}_2 \qquad [32]$$

Figures 11 and 12 show the normalized constants $\hat{\xi}_1$ and $\hat{\xi}_2$ as a function of the geometric mean of $z_a$. From Figure 11, we observe that as the geometric mean of $z_a$ increases, i.e. the adhesive force holding the particles on the surface increases, the value of normalized constant $\hat{\xi}_1$ in the modified model can reach as much as twice that in the original model for the same spread factor. Figure 12 shows that the normalized constant $\hat{\xi}_2$ for both the original and modified is very close to 1, In particular as the geometric mean of $z_a$ increases the value of $\hat{\xi}_2$ decreases to around 1.06 for the modified model and 1.055 for the original model, regardless of the spread factor. Because of this, from Eq.[32], we know that the normalized long term resuspension rate has a very small dependence on $\omega$, since the power of $\omega$, namely ($\xi_2$ - 1) is very close to zero.

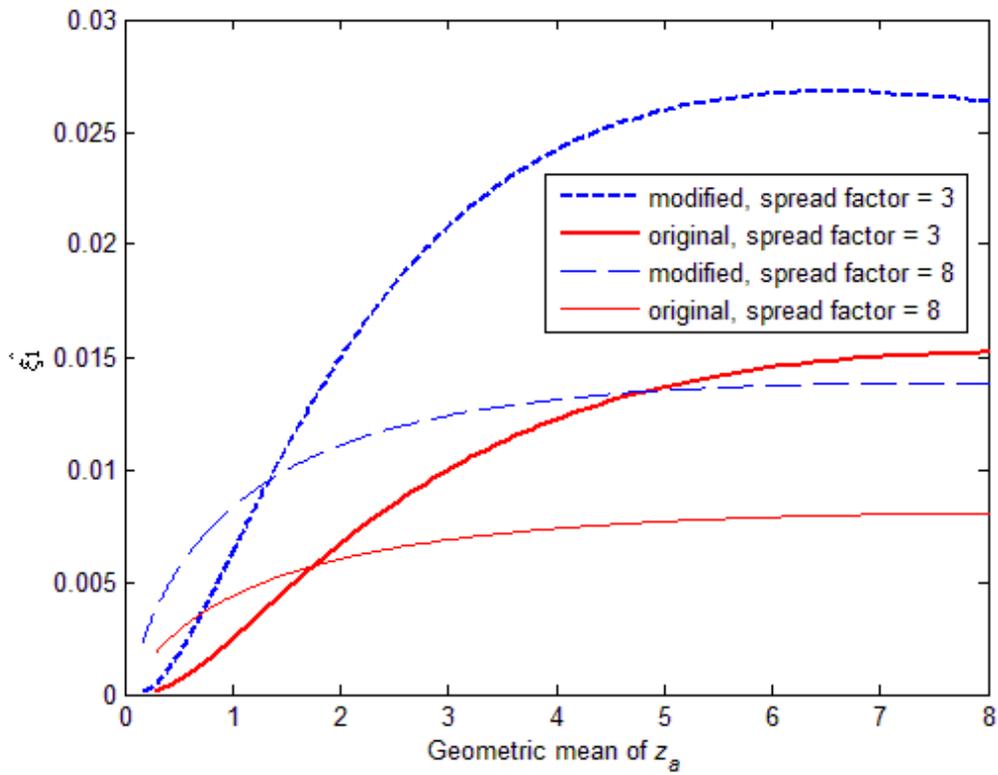

**Figure 11** – constant $\hat{\xi}_1$ for normalized long term resuspension rate $\Lambda/\omega$ vs. geometric mean of normalized adhesive force $z_a$

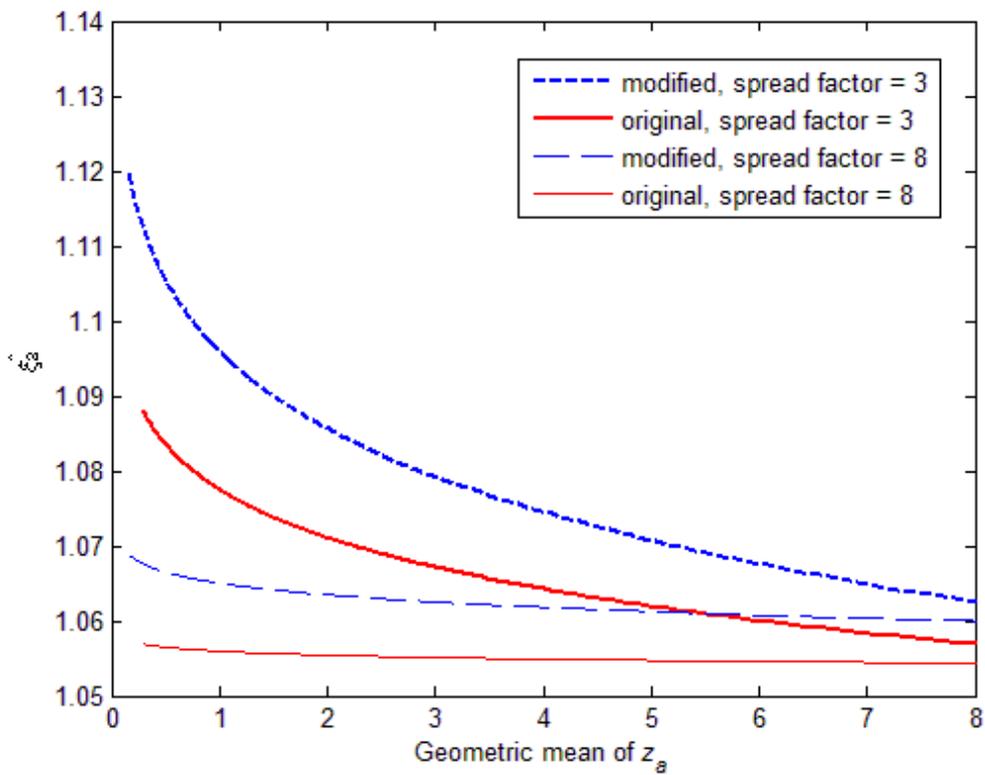

**Figure 12** - Inverse power for long term resuspension rate $\Lambda/\omega$ vs. geometric mean of normalized adhesive force, $z_a$

### 5.3 Comparison of Original and Modified RnR model

There are several points that need to be emphasised before we make a comparison of the predictions made by the two models.

- O'Neill's formula (Eq.[15]) is used to calculate the resultant fluctuating aerodynamic from the fluctuating streamwise velocity.
- The parameters from the DNS data at $y^+ = 0.1$ are used in the modified model because the value of $y^+$ is much closer to the value of the typical particle radius $r^+$ (in wall units) than the other values of y+ (See Table 2). Although it is shown in Table 2 that the typical burst frequency $\omega^+$ varies with $y^+$, at the moment the typical burst frequency $\omega^+$ value is a fixed value chosen from the case $y^+ = 0.1$ due to the fact that there are not enough simulation data to produce the relationship between $\omega^+$ and $y^+$. This will be recommended in future work.
- For Hall's experimental data Biasi's correlation (Eq.[7]) is applied to both the modified and original models to calculate the reduction and spread in adhesion as a function of particle size (since the correlation itself is based on Hall's data). The removal fraction measured in the Ibrahim *et al.* experiment suggests a much narrower spread in adhesion than measured in Halls experiment and in this case we have based the spread on their surface roughness measurements.

#### 5.3.1 Hall's resuspension and adhesion measurements

We recall that in Hall's experiment (Reeks & Hall, 2001) there were three types of particles (10μm alumina, 20μm alumina and 10μm graphite) used in the experiment. Hall measured both the adhesive force and resuspension of those particles.

There were 20 resuspension runs for both graphite and alumina particles performed in the experiment. Here the experimental data of Run – 9, 10, 15 (for 10μm alumina, in diameter) and Run – 7, 8, 20 (for 20μm alumina particles) will be used to compare experimental results for the fraction resuspended the modified and original model predictions.

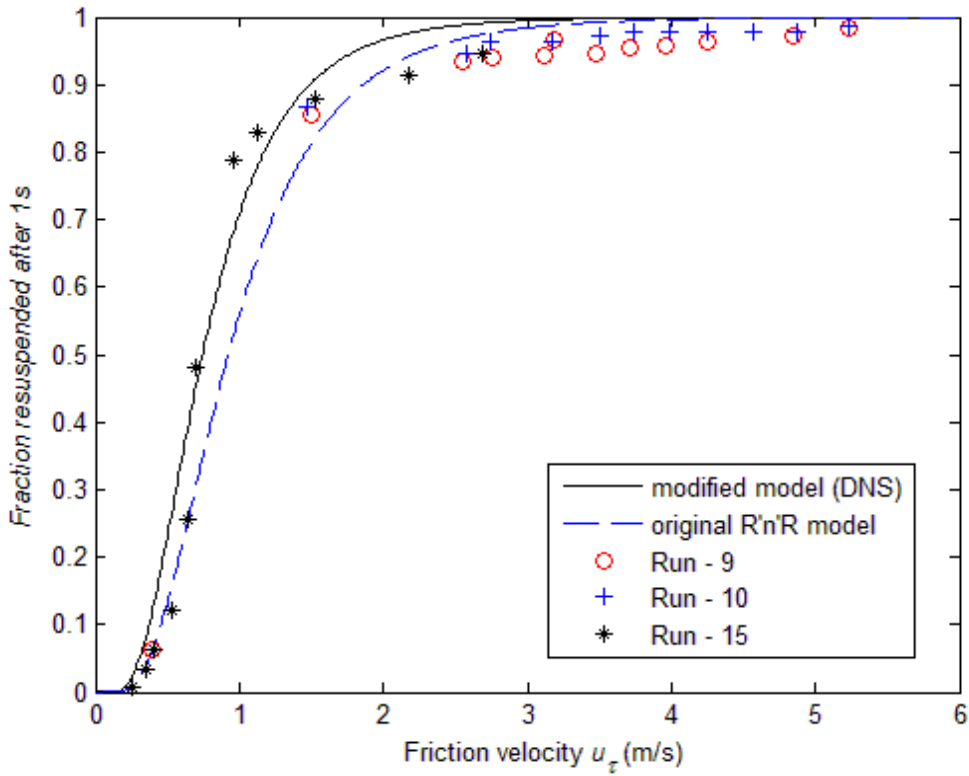

**Figure 13** - Comparison of measurements resuspension fraction of 10 micron alumina particles in Hall's experiment with model predictions

Note the calculation of the fraction resuspended after 1 s is a nominal time, just long enough for this time to be sufficient for the resuspension rates at the end of the exposure time to be very small (and in the long-term resuspension range).

Figure 13 shows the comparison of resuspension fractions calculated using the modified and original RnR models with the experimental data for 10μm alumina particles. It can be observed that the modified model gives results closer to the experimental data around the friction velocity for 50% removal, where the resuspension is most sensitive to changes in the flow. Although the modified model gives more resuspension than the experimental data when friction velocity is smaller than 0.5m/s and larger than 1.5m/s, the solid curve still on the whole gives better agreement with the experimental data than the original model. This observation is also true for 20μm alumina particles as can be observed in Figure 14. However it is to be noted that there is considerable scatter in the experimental results resulting from errors in both the absolute measurement of the adhesive forces and in the resuspension itself, whilst the very broad spread in adhesion tends to obscure any fundamental differences between the original and modified model predictions. This in itself may be regarded as a useful result. In most measurements of resuspension the absolute value of adhesive forces and their spread are unknown or subject to a large degree of uncertainly. What these results and comparison show is that even when there is a significant difference in the values of the model parameters as is the case here, because of the broad spread in adhesion normally present, this does not have a marked difference on the predicted levels of resuspended fraction. However this is not the case with the long term resuspension rates as we have shown previously (see Figures 9 and 10) which unfortunately were not measured in this experiment.

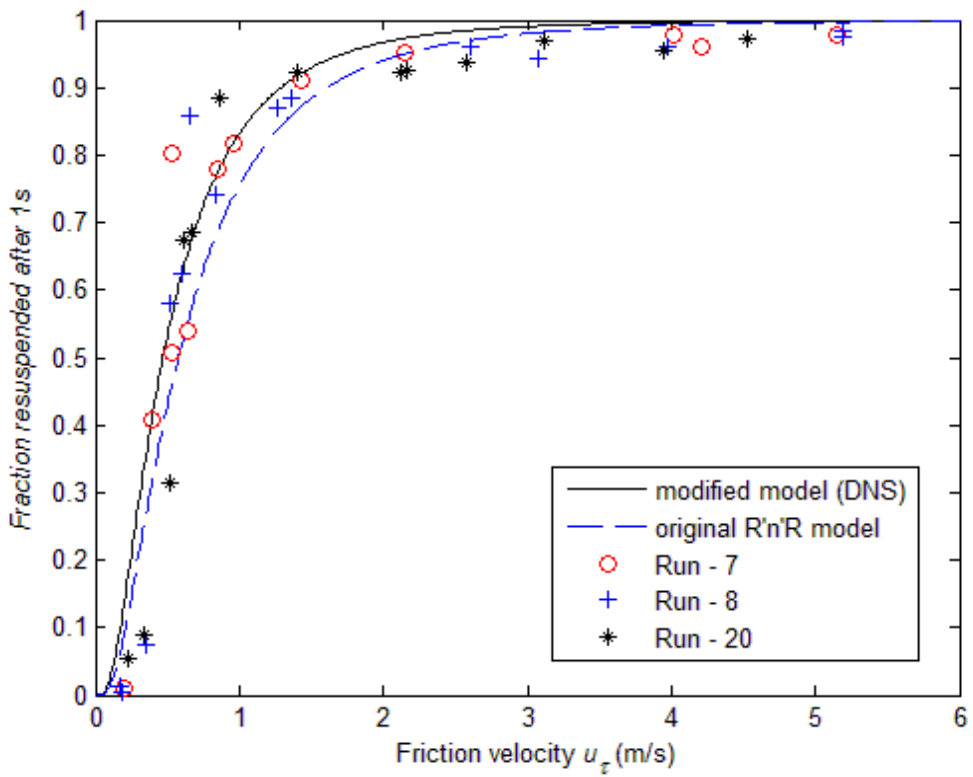

**Figure 14** - Comparison of measurements resuspension fraction of 20 micron alumina particles in Hall's experiment with model predictions

To investigate the difference between the modified and original model predictions, the effect of two important parameters (the typical burst frequency $\omega^+$ and the rms coefficient $f_{rms}$) are studied here. The Table 3 shown below highlights the differences in these two parameters.

|  | $\omega^+$ | $f_{rms}$ |
|---|---|---|
| Modified (DNS) | 0.1642 | 0.366 |
| Original | 0.0413 | 0.2 |

**Table 3 -** Values of $\omega^+$ and $f_{rms}$ used in modified and original model

We have calculated the fraction resuspended after 1s as a function of friction velocity, and resuspension rate as a function of time for the monodisperse 10μm alumina particles using a reduction in adhesion of 0.0105 and a spread in adhesion of 3.095 based on Biasi correlation (Eq.[7]). Hall's experimental conditions are used as the basis of this exercise.

| Fluid density ($kg.m^{-3}$) | Fluid kinematic viscosity ($m^2.s^{-1}$) | Surface energy ($J.m^{-2}$) |
|---|---|---|
| 1.181 | 1.539 x $10^{-5}$ | 0.56 |

**Table 4 -** Parameters of Hall's experimental conditions

The results based on the parameters in Table 4 are shown below.

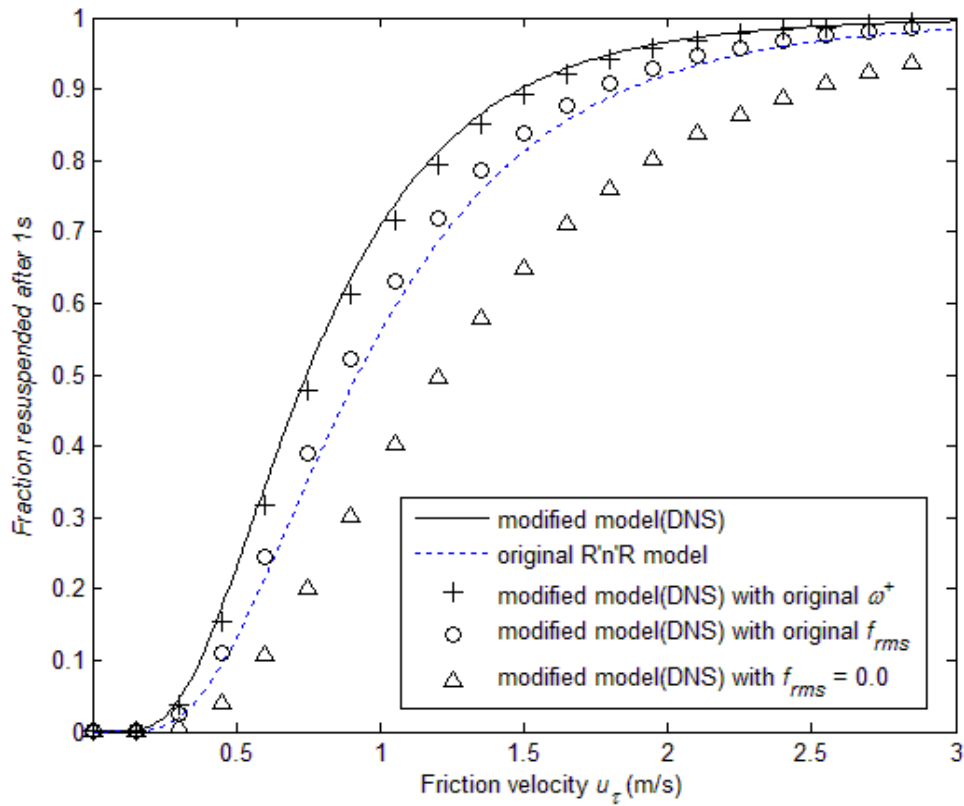

**Figure 15 -** Comparison of resuspension fraction after 1s between modified and original models using Hall's experimental flow and adhesion properties for 10 micron alumina particles

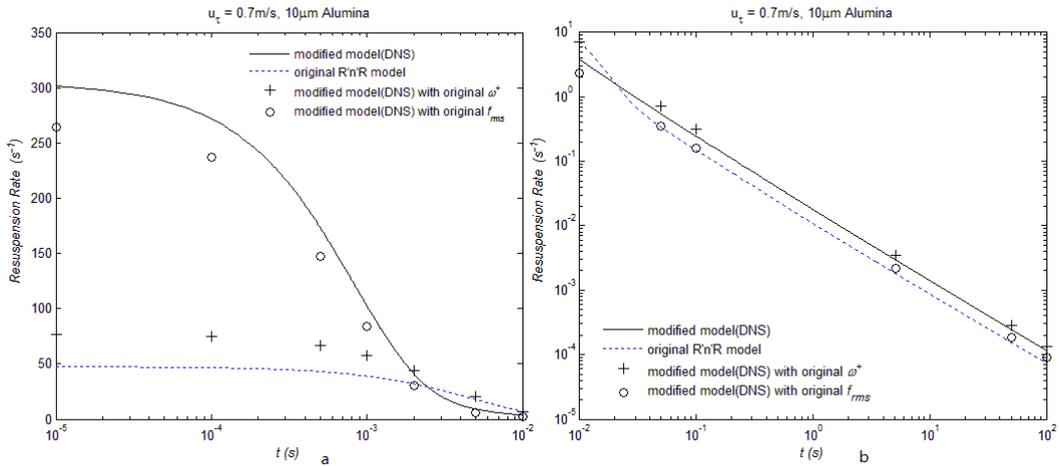

**Figure 16** - Comparison of fractional resuspension rates for modified and original models

Figures 15 and 16 show the influence of changing the values of ω and $f_{rms}$ in the modified model to their original values upon the fraction resuspended after 1s (Figure15) and the resuspension rate for a friction velocity of 0.7m/s (Figure 16) corresponding to 50% removal predicted by the modified model from Figure 15. To show the influence of the turbulence on the resuspension, Figure 15 also shows the fraction resuspended with $f_{rms}$ set to zero. That is the resuspension in this case is entirely due to a threshold based on a balance between the adhesive fore and the mean removal force.

With reference to these two Figures, there are three points to be noted:

1. The effect of changing ω on the resuspension fraction after 1s is hardly noticeable even though the modified value of ω is 4 × the original value. This is because the nominal exposure time of 1 s is well outside the range for short term resuspension (~$10^{-2}$ s from Figure 16b) so all of the fraction of particles on the surfaces associated with short term resuspension has been removed (more precisely $ωt > 25$ from Figure 10). However in contrast, the value of ω affects dramatically the short term resuspension rate for $ωt <~ 1$ (which in Hall's experiment corresponds to $t < 0.1$ msec): in particular the initial resuspension rates for the modified model are a factor of 6 × that given by the original model, this difference being due to a factor of 4 in the values of ω and a factor of 1.5 due to the influence of non-Gaussian over Gaussian distributions for $\dot{f}$

2. The rms/mean removal force ratio, $f_{rms}$ is the most influential parameter for the fraction removed over the total short term period. As we observe from Figure 15, when the original value of the rms coefficient $f_{rms}$ (0.2) was used in the modified model (circle symbol) the result is much closer to the original model result (dotted line). The small but noticeable difference in this case is due to the non-Gaussian distribution as opposed to the Gaussian distribution for the removal forces.

3. Whilst there is a factor of 6 in the ratio of the modified over the original model values for the initial resuspension rates which mostly reflects the significant difference in the values of ω, we note from Figure 16a that beyond an exposure time ~ $2 \times 10^{-3}$ s the resuspension rates predicted by the original model with the lower value of ω are greater than that predicted by the modified model, with the result that the total integrated removal fraction over the total short term period ~$10^{-2}$s, is only weakly dependent on ω (to be consistent with the lack of dependence on ω of the fraction removed shown in Figure 15). This implies that the increased fraction removed in the very short term period ($<~ 2 \times 10^{-3}$ s ) is compensated by the reduced fraction removed during the period ( $2 \times 10^{-3}$ s $<~10^{-2}$ s ). We note that the difference in the long term resuspension rates (shown in Figure 16b) reflects the dependence on $f_{rms}$ and to a lesser extent the difference between the Gaussian and non Gaussian distributions for $f$ and $\dot{f}$ (see Figure 3)

Finally we note that the difference between the removal fraction for modified and original models could become significant when the friction velocity is small. As shown in Figure 17, the ratio increases to around 6 or 7 on a nominally very smooth surfaces (spread = 2) when the friction velocity is smaller than 1m/s.

**5.3.2 Ibrahim *et al.*(2003) measurements of surface roughness and particle resuspension**

Ibrahim *et al.* (2003) conducted a series of experiments to characterise the detachment of micro-particles from surfaces exposed to turbulent air flow during an accelerated free stream flow. Smooth glass plates used as substrates were scanned with an atomic force microscope to determine their roughness height distribution and converted into values for the adhesion reduction factor using the results of Cheng *et al.* (2002). Micro particles of different sizes and materials and shapes (mostly micro−spheres) were deposited as sparse monolayers onto the substrates under controlled clean and dry conditions. Micro video graphic observations of individual micro-particle detachment showed that detachment occurred primarily by rolling and not lift –off although entrainment (resuspension) did not always occur following rolling.

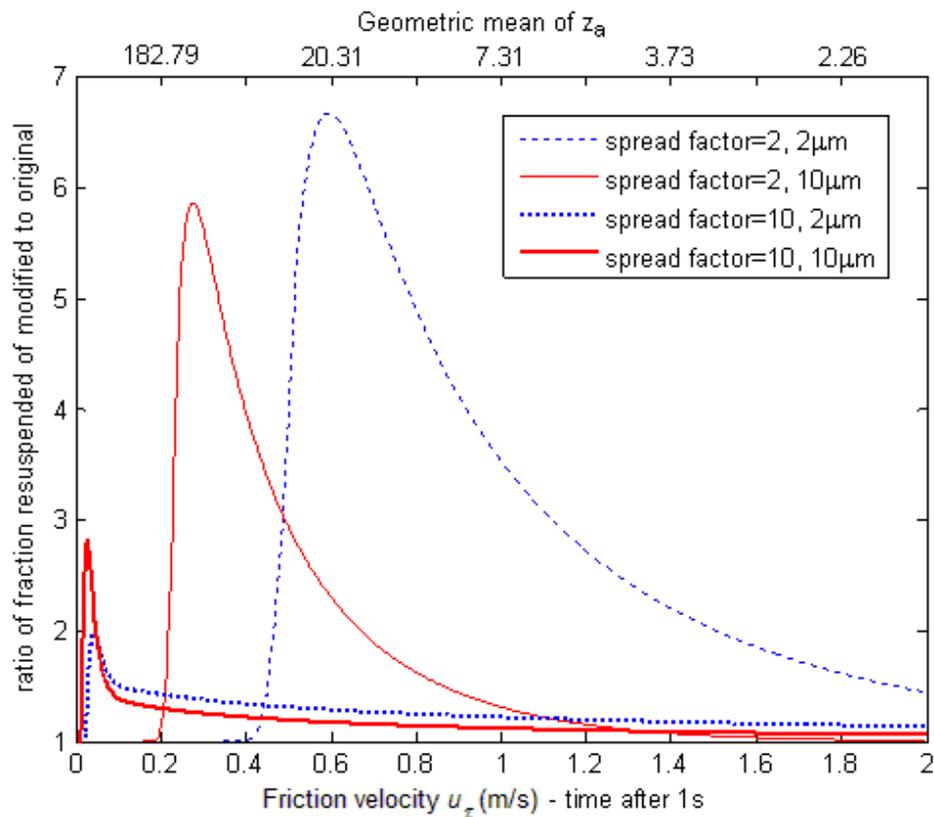

**Figure 17** - Comparison of resuspension fraction ratio of modified (DNS) model with original model for Hall's experimental conditions

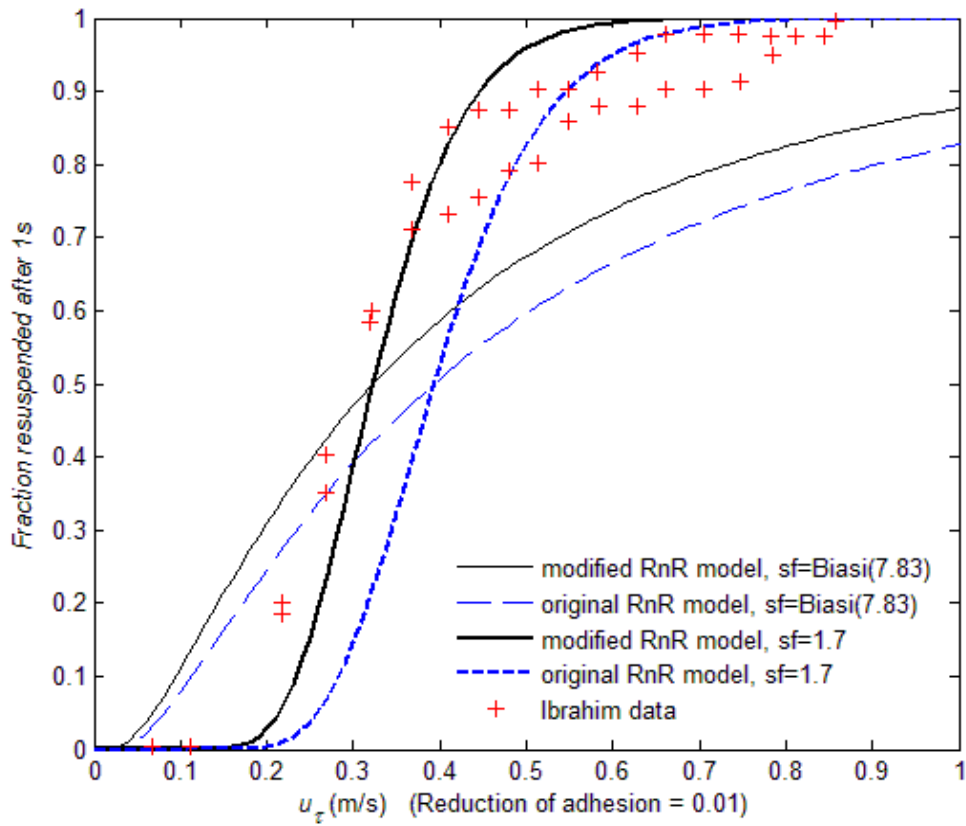

**Figure 18** - Comparison of modified (DNS) RnR model and original RnR model predictions with experimental results of Ibrahim *et al.* (2003) for the resuspended fraction of 30 micron lycopodium spores on glass substrate

Measurements were made of the detachment during an accelerating phase which occurred over a linear change in the free stream velocity over a period of ~ 60 -150 s. The timescale for fluid motion in a fully developed turbulent boundary layers ~ msec so the flow during this phase can be assumed quasi-steady. As an example for comparison we have chosen the results for the removal fraction of lycopodium spores as a function of friction velocity which exhibits the narrowest range of friction velocities for the size of particles considered. The range is much narrower than that observed in Hall's experiment and would correspond to a very narrow spread in the surface adhesive forces reflecting the small value of 1.7 for the geometric spread in the rms of roughness height.

This is the ideal case to show up differences between the modified and original models. The original and modified model predictions are shown in Figure 18, based on a lycopodium/glass interfacial surface energy of 0.3 J/m$^2$, a reduction in adhesion of 0.01 (based on Cheng *et al.* (2002) measurements) and a geometric spread in adhesion of 1.7 with a value of $r/a$ =100 based on Hall's measurements (Reeks & Hall 2001). There is a marked difference between the non-Gaussian model and original Gaussian model which in this case is due to the significant difference (a factor of almost 2) in the ratio of the rms / mean of the drag force $f_{rms}$. We note for instance that the friction velocity (~0.35m/s) for removal of 50% particulate based on the modified resuspension model, removes ~30% based on the original model. The friction velocity required to remove 50% in the original model is increased by 20% to 0.4 m/s. Also on the same graph is shown the comparison between the two model predictions using the same value for the reduction in adhesion and a value of 7.5 for the spread in adhesion based on Biasi's correlation. We note that the range of friction velocities for removal is greatly increased and the difference between the modified and original model much diminished.

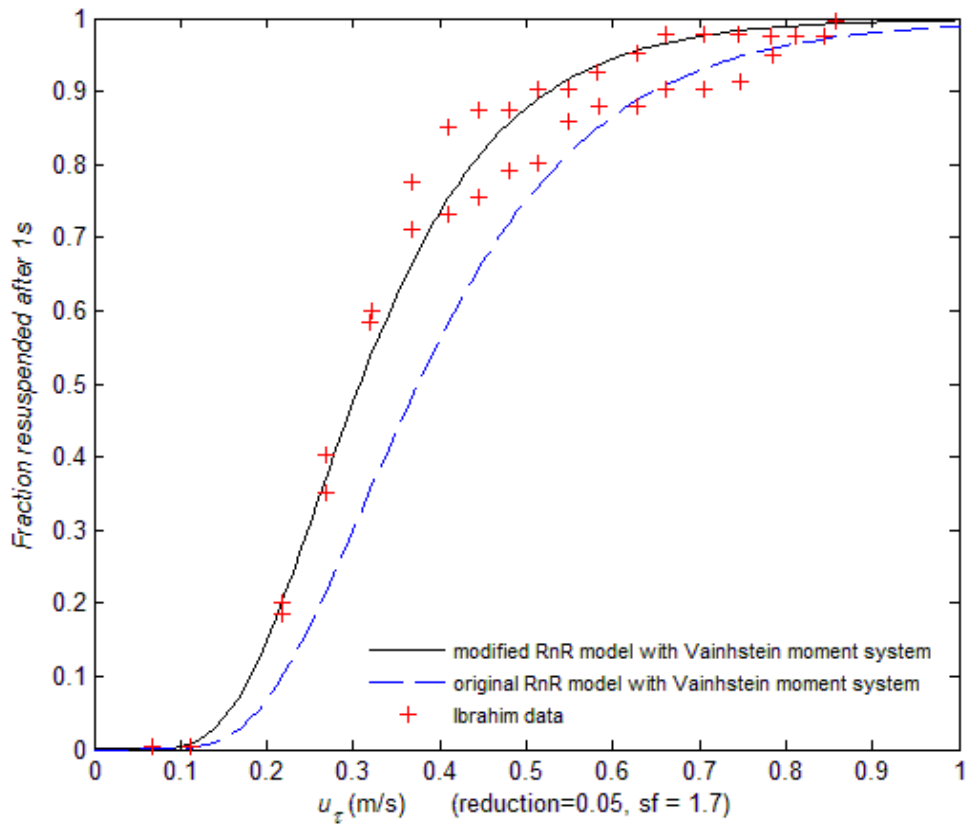

**Figure** 19 – Comparison of Vainhstein *et al.* model predictions with Ibrahim et al (2003) experimental results for the resuspended fraction of 30 micron lycopodium spores on a glass substrate

Figure 19 shows the modified and original models' predictions based on the Vainhstein *et al.* system of adhesion and drag moments about a single asperity. Zhang (2011) showed that for particles > 6 microns in size the Vainhstein *et al.* model gives greater removal than for the equivalent RnR model (based on the balance of the moments of the mean drag with the adhesive force). This is reflected in the removal fraction where we have used the same parameter values as in the RnR model except for the reduction in adhesion which we have increased to 0.05 to give good agreement using the modified non-Gaussian model. As we might have expected the difference between the modified and original model predictions is the same as that for the RnR models.

**5.3.3 Sensitivity to material properties and surface roughness**

The removal fraction depends upon a number of material/surface and fluid properties of which the most uncertain are associated with the particle surface interactions. We have seen how much surface roughness influences the adhesion – even for nano-size roughness heights, noting that roughness influences not only the adhesive forces but also the adhesive moments. In the Vainhstein *et al.* model the adhesive moment depends on the radius of the contact area of a surface asperity which in turn depends upon its radius of curvature; in the Reeks & Hall RnR model it depends upon the distance between asperities. Methods exist for calculating the adhesive force from the roughness geometry as has been used by Ibrahim et al (2003) and in Rabinovich *et al.* (2000a,b). It is useful to indicate how sensitive the removal fraction predictions are to the values of the various material constants and roughness parameters. This shown in Table 5 for both versions of RnR model.

| **Modified RnR Model** | **Fraction Resuspended (after 1s)** (Friction velocity = 0.35 m/s) |
|---|---|

| Parameter name | Base Value | Base Case | Parameter Doubled | Parameter Halved |
|---|---|---|---|---|
| Ratio of Particle ($r$) and Asperities distance ($a$) | 100 | 0.626 | 0.947 | 0.166 |
| Surface Energy, J/m$^2$ | 0.3 | 0.626 | 0.166 | 0.947 |
| Reduction in Adhesion | 0.01 | 0.626 | 0.166 | 0.947 |
| Spread of Adhesion | 1.7 | 0.626 | 0.556 | - |
| **Modified RnR Model with Vainhstein *et al.* Moment System** | | **Fraction Resuspended (after 1s)** (Friction velocity = 0.1 m/s) | | |
| Parameter name | Base Value | Base Case | Parameter Doubled | Parameter Halved |
| K, Pa | 5.76 x 10$^{10}$ | 0.613 | 0.708 | 0.512 |
| Surface Energy, J/m$^2$ | 0.3 | 0.613 | 0.227 | 0.907 |
| Reduction in Adhesion | 0.01 | 0.613 | 0.157 | 0.943 |
| Spread of Adhesion | 1.7 | 0.613 | 0.550 | - |

**Table 5** - Parameter sensitivities on resuspension fraction of modified non-Gaussian versions of RnR model

## 6. A comparison with model predictions based on Lee & Balachandar's measurements of the drag force

In the previous studies, the drag force acting on a particle was calculated using the modified Stokes drag formula given by O'Neill (1968) for the drag force of a spherical particle on or near a wall. Recently Lee & Balachandar (2010) have made extensive calculations of the aerodynamic forces acting on a small particle on or near a wall in a turbulent boundary layer generated by DNS. In what follows we shall use these results to calculate the corresponding drag forces generated in our DNS flow and compare the model predictions of the resuspension with those based on O'Neill's formula.

*Application of O'Neill's Formula*
Assuming the local fluid velocity is similar to the particle velocity, the instantaneous drag force acting on a spherical particle is then calculated from the velocity by applying O'Neill's (1968) formula.

$$F = 1.7 \cdot 6\pi\mu_f r u$$

where $r$ is the particle radius and represents the distance of the particle from the wall (i.e. corresponding to $y^+$). The fluctuating drag force $f$ is obtained by subtracting the mean $<F>$ from $F$ (i.e., $F - <F>$) and then normalized by its rms value, namely

$$z_S = \frac{f}{\sqrt{\langle f^2 \rangle}} \qquad [33]$$

The first derivative of fluctuating force is calculated via $\overset{\&}{f} = \frac{f_{i+1} - f_i}{\Delta t}$, and then normalized as

$$\dot{z}_S = \frac{\dot{F}}{\sqrt{\langle \dot{F}^2 \rangle}} \qquad [34]$$

where $z_S$ is the normalized fluctuating drag force and $\dot{z}_S$ is its first derivative.

### *Application of the Lee & Balachandar (2010) Results*

Lee and Balachandar (2010) (L&B) calculated numerically the contributions to drag and lift on a spherical particle arising from shear, translation and rotation mechanisms applicable at modest Reynolds numbers. Here the particle is considered as sitting on the wall and the lift force is neglected. Therefore, the translation and rotation force are not considered. The drag force is expressed in the form:

$$F = C_D \cdot \frac{\pi}{2} \rho_f G |G| L_W^2 r^2 \qquad [35]$$

where $C_D$ is the drag coefficient solely due to the local shear, $G$ is the local shear rate $L_w$ is the distance from the wall to the centre of the particle and $r$ is the particle radius . The drag coefficient is given by

$$C_D = \frac{40.81}{Re_r}\left(1 + 0.104 Re_r^{0.753}\right) \qquad [36]$$

where $Re_r$ is the shear Reynolds number which is given by

$$Re_r = \frac{2|G|L_W r}{\nu_f} \qquad [37]$$

The distance from the wall to the centre of the particle can be written as

$$L_W = \frac{\nu_f y^+}{u_\tau} \qquad [38]$$

which in this case is the same as the particle radius $r$.

From the DNS data, we obtained the instantaneous velocity gradient (dU/dy) for a given $y^+$ (e.g., $y^+ = 0.1$). Figure 20 shows the best-fit Rayleigh disributions for the normalised drag force and based on the O'Neill and L&B formulae. We note there is a noticeable but not appreciable difference in around the peak values of the distributions for both variables: the variations resulting from the different values of $y^+$ are in fact more significant. Similar results were obtained for the distribution of the normalised time derivative (see Zhang 2011) based on the Johnson SU distribution.

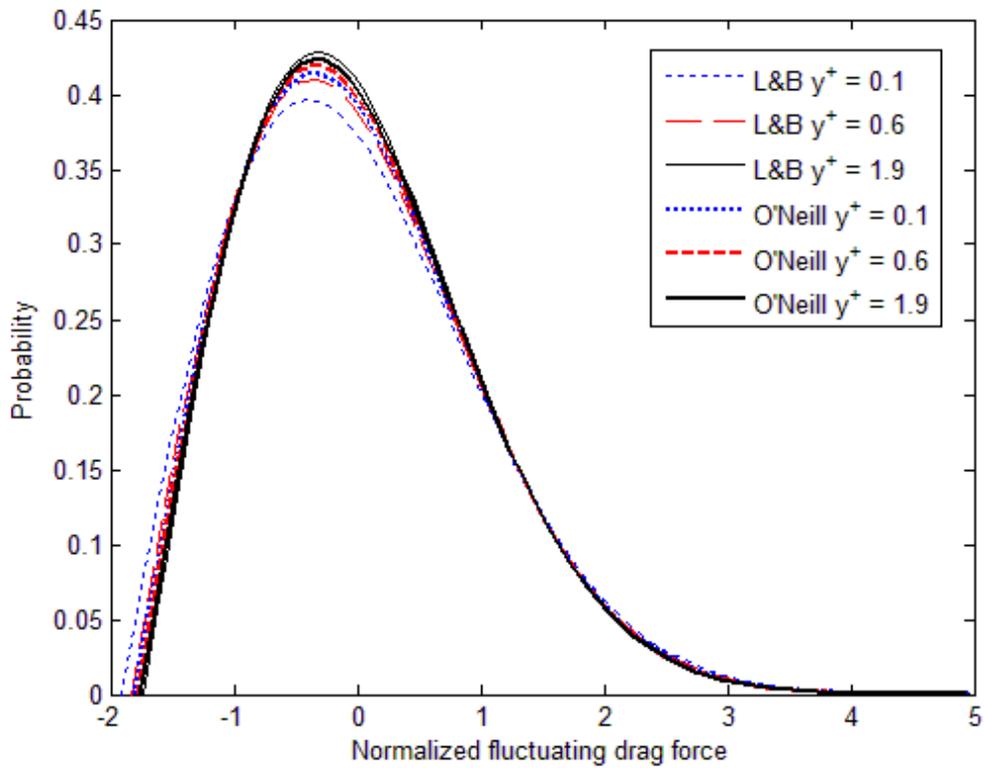

**Figure 20** - Distribution of normalized fluctuating removal force based on O'Neill's and L&B formulae

Following the steps Eq.[19] to Eq.[22], the parameters are listed below with the values based on O'Neill's and the L&B formulae for the drag force.

| O'Neill formula | $B_{f_k}$ | $A_1$ | $A_2$ | $\omega^+$ | $f_{rms}$ |
|---|---|---|---|---|---|
| $y^+ = 0.1$ | 0.3437 | 1.8126 | 1.4638 | 0.1642 | 0.366 |
| $y^+ = 0.6$ | 0.3469 | 1.7848 | 1.4466 | 0.1520 | 0.366 |
| $y^+ = 1.9$ | 0.3512 | 1.7599 | 1.4313 | 0.1313 | 0.365 |
| L&B formula | $B_{f_k}$ | $A_1$ | $A_2$ | $\omega^+$ | $f_{rms}$ |
| $y^+ = 0.1$ | 0.3699 | 1.9179 | 1.5295 | 0.1372 | 0.346 |
| $y^+ = 0.6$ | 0.3621 | 1.8364 | 1.4786 | 0.1276 | 0.370 |
| $y^+ = 1.9$ | 0.3498 | 1.7317 | 1.4140 | 0.1293 | 0.447 |

**Table 5** - Comparison of parameters by two formulae of calculating fluctuating force

From the Table above, we observe that the parameters calculated by these two formulas are on the whole significantly similar except for the value of $f_{rms}$ for $y^+ = 1.9$. It is noted that unlike the application of the O'Neill formula, the rms coefficient $f_{rms}$ using the L&B formula increased with increasing $y^+$

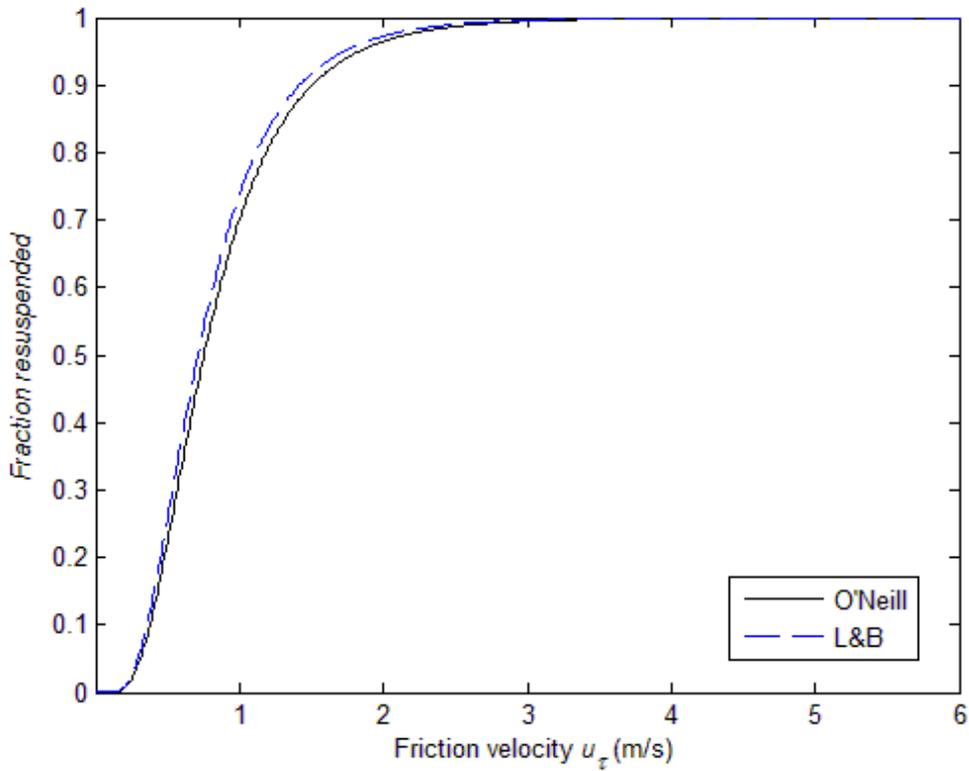

**Figure 21** - Comparison of resuspension fraction of the two statistics-generation formulae ($y^+$ = 1.9) using Hall's experimental flow and adhesion properties for 10 micron alumina particles

The comparison shown in Figure 21 based on the DNS data for $y^+$ = 1.9 for the fraction resuspended fraction indicates that the predictions calculated from the L&B formula are quite close to the original model using O'Neill's formulae for the spread in adhesion typical of those observed in Halls experiments (less than 5%). The difference is more marked for a very narrow spread (e.g. spread factor = 1.5) and reflected in the difference in values of $f_{rms}$ especially in the case of $y^+$=1.9. For the reason explained previously the difference in values of $\omega^+$ has little effect on the long term resuspension predictions. On the grounds of the simplicity of application of O'Neill's formula, we will use this formula in our modified RnR model in subsequent analysis of resuspension from multilayer deposits. For more precise details see Zhang (2011).

### 7. Summary and Discussion

This paper has considered the influence of non-Gaussian aerodynamic forces and moments on the removal of micron size particles from a surface exposed to a turbulent boundary layer. The focus has been on sparse monolayer surface coverages so that any individual particle maybe considered in isolation from its neighbours. We have chosen to consider the influence of these aerodynamic forces on the RnR kinetic model for particle resuspension which is based on an evaluation of the resuspension rate constant for a particle rocking and rolling about the surface roughness asperities between particle and substrate. This mechanism is consistent with experimental observations which indicate that incipient rolling rather than lift is the principal mechanism for particle removal from a surface. Whilst other more detailed models might have been considered e.g. the Monte Carlo approach of Guigno and Minier (2008), the RnR model was chosen because of its adaptability and computational efficiency. We recall that the formula for the resuspension rate constant in the original RnR was evaluated using Gaussian statistics, though the general formula is appropriate for any form of statistics. It is natural therefore to compare the predictions of the original RnR model with those of the modified (non-Gaussian) version of the model.

The formula for the resuspension rate constant depends upon the distribution of the aerodynamic moments and their time derivatives. In the RnR model as in other models e.g. that of Vainhstein et al. (1997), the moment of the drag force is the principal moment responsible for removal. This is converted to an equivalent fluctuating removal force $f$ and its time derivative $\dot{f}$ which shows the drag forces amplified by a factor of $r/a$ where $r$ is the particle radius and $a$ an effective distance between asperities (in the RnR model $r/a \sim 100$ based on experimental measurements). The values of $f$ and $\dot{f}$ and their probability distributions are obtained from the calculation of the streamwise velocities in a DNS of a fully developed turbulent boundary layer which are then converted to aerodynamic forces using O'Neill's formula and also for comparison with the more recent and more accurate DNS based formulae of Lee and Balanchandra (2010) calculations. These show the distributions to be strongly non-Gaussian, best fitted by a Rayleigh distribution for $f$ and a Johnson SU distribution for $\dot{f}$.

Based on these DNS results, the resuspension rate constant is modified in three separate ways from its original value. Firstly through the distributions of $f$ and $\dot{f}$ (normalised on their rms values), secondly through the typical removal frequency of particles $\omega = (\langle \dot{f}^2 \rangle / \langle f^2 \rangle)^{1/2}$ from a surface and thirdly through the ratio of the rms /mean of the removal force, $f_{rms}$. We examined the influence of all three independently upon the fraction resuspended and the short and long term resuspension rates and then all together when we compared modified and original model predictions with experimental results for the fraction resuspended.

**7.1 Dependence on Gaussian versus non-Gaussian distributions for $f$ and $\dot{f}$**

In the formula for the resuspension rate constant, the probability distribution of the fluctuating removal force $F$ determines the particle concentration on the surface at the detachment point when $f = f_a - \langle F \rangle$. It was noted that when the adhesive force / rms removal force ratio $z_a$, is large ($z_a \sim 8$), the ratio of the resuspension rate constant based on the non-Gaussian to that of the Gaussian model is $\sim 30$, reflecting the much slower decay of the non-Gaussian Rayleigh distribution for the aerodynamic drag force in the tails of the distribution compared to that of the Gaussian distribution. However the broad range of adhesive forces in practice significantly reduces the influence of the tails of the distribution mainly because the contribution to the resuspension in this region of the adhesive force distribution is so small. We note from Figure 7 that for very large values of the spread in adhesion $\sim 10$, the ratio of the initial resuspension rates approaches an asymptotic value $\sim 1.5$. This value is also typical of the maximum value of ratio of the long term resuspension rates $\sim 1.3$ shown in (Figure 8) for various adhesive spread factors. As for the influence of the distribution of $\dot{f}$, we can see from the general formula for the resuspension rate constant in Eq.(10) that this is reflected in the net value of $\dot{f}$ for $\dot{f} \geq 0$. For a Gaussian it is $\langle \dot{f} \rangle^{1/2} / 2\pi \sim 0.16 \langle \dot{f} \rangle^{1/2}$ and for the non-Gaussian it is $B_{\dot{f}} \langle \dot{f} \rangle^{1/2} \sim 0.35 \langle \dot{f} \rangle^{1/2}$ (see Table 2) where the increased contribution arises from the significant difference in the tails of the distribution for $\dot{f}$ (Figure 3).

**7.2 Dependence on the removal frequency $\omega$ and rms/ mean removal force $f_{rms}$**

The removal frequency $\omega = \sqrt{\langle \dot{f}^2 \rangle / \langle f^2 \rangle}$ is the natural removal frequency associated with the fluctuating removal force $f(t)$, being the removal frequency for a uniform distribution of both $f$ and $\dot{f}$. We note that it is weighted towards the higher frequencies of the removal force (and by implication those of the streamwise velocity fluctuations) and that it is intrinsically higher than the integral time scale of $f(t)$. Its DNS value is a factor of 4 greater than the corresponding value used in the original model based on the experimental measurements of Hall of the fluctuating lift force (Reeks & Hall 2001). The resuspension rate scales on $\omega$ whilst the exposure time scales on $\omega^{-1}$, so that in real time, as the resuspension rate decays so these scaling factors oppose one another in their influence on the resuspension rate. This is most noticeable in the long term when the long term resuspension rate is almost independent of $\omega$ (see Eq. (30). What is noticeable is that although the resuspension rate in the short term varies significantly with $\omega$, so long as the exposure time extends into the long term resuspension rate ($\omega t > 30$, see Figure 9) where the resuspension rate is very low compared with the initial rate, then the integrated removal is almost independent of $\omega$ and insensitive to the actual exposure time. Hence the requirement to quote only a nominal exposure time of time of 1s when evaluating the removal fraction in this range. This is of course the case for most experiments where the removal fraction is measured.

Whilst the non-Gaussian distribution of $f$ in the modified models increases the fraction resuspended in the short term period, the difference in the values of the coefficient $f_{rms}$ has the most significant effect (see Figure 15). This is also true of the long term resuspension rates predictions. We recall that the long term resuspension rate was given by $\Lambda(t) = \xi_1 t^{-\xi_2}$ and we evaluated the coefficients $\xi_1$ and $\xi_2$ as a function of the geometric mean and spread of the normalised adhesive force $z_a$, see Figures 11 and 12, where because of the weak dependence on $\omega$, the values of the coefficient $\xi_1$ are almost the same as those of $\hat{\xi}_1$ and $\xi_2 = \hat{\xi}_2$. From Figure 11, we observe that as the geometric mean of $z_a$ increases, i.e. the adhesive force holding the particles on the surface increases, the value of $\xi_1$ in the modified model can reach as much as twice that in the original model for the same spread factor. From Figure 15 we recall that $\xi_2$ is close to 1, but that the value of $\xi_2$ for the modified model is consistently greater than that for the original model.

**7.3 Comparison of modified and original RnR models with experimental results**
We compared original and modified model predictions for the fraction of particles resuspended with the experimental results of Hall (Reeks and Hall, 2001) and Ibrahim *et al.* (2003). The experimental measurements of Hall had been used before to validate the original RnR model (where we note that Hall used separate centrifuge measurements to obtain the reduction and spread in the adhesion and the drag force amplification factor *r/a*). On the whole the modified model predictions gave marginally better agreement with the experimental data than the original model in the most sensitive part of the resuspension curve although we noted that that there is considerable scatter in the experimental results whilst the very broad spread in adhesion observed tends to obscure any fundamental differences between the two model predictions. Nevertheless we regarded this to be a useful result, pointing out that even when there is a significant difference in the values of the model parameters as is the case here, because of the broad spread in adhesion normally present, this does not have a marked difference on the predicted levels of resuspended fractions.

The noticeable feature of the Ibrahim *et al.* measurements of the resuspension was that they corresponded to a narrow range of adhesive force consistent with measurements of the surface roughness measurements from which the reduction in adhesion was evaluated. Absolute values of the adhesion depended upon values for the surface energy of the particle and surface materials of which there is much uncertainty with only nominal values given. The absence of a large spread in adhesion as in Hall's measurements meant a greater sensitivity to the influence of non-Gaussian forces model and ratio of the rms to mean removal force, $f_{rms}$. However the comparison with experimental measurements should be regarded as a bench marking exercise since no measurements were made of the amplification factor *r/a* and there is no guarantee that it is the same value as that measured in Hall's experiment (and whose value was used to obtain the RnR predictions shown in Figure 18). We also compared the predictions of modified non-Gaussian and original Gaussian versions of the Vainhstein *et al.* model with the experimental resuspension data where a good fit to the data for the modified version was obtained using a value of 0.05 for the adhesion reduction instead of 0.01 as used in RnR predictions: this indicates that the Vainshstein *et al.* model in this case predicts more resuspension than the RnR model using the same model parameters (note the value of *r/a* is not an independent parameter in the Vainhstein *et al.* model)

**7.4 Use of Lee & Balachandar's drag force formula**

Finally, we have examined the implications for resuspension of using the recently published formula of Lee & Balachandar(L&B) (2010) for the drag force acting on a particle on a surface (based on their DNS data of drag forces on particles on or near a surface in a turbulent boundary layer); the original RnR model uses O'Neill's formula. The comparison indicated that the resuspension predictions using the L&B formula were quite close to those of the original model for the drag force except in the region $-1 < z_1, z_2 < 1$ ($z_1$ and $z_2$ are drag force and its derivative normalized on their rms) where use of O'Neill's formula gives higher values. However, this has very little effect on the resuspension rate and resuspension fraction. Therefore, on the grounds of the simplicity of application of O'Neill's formula, we will use this formula in our modified RnR model in subsequent analysis of resuspension from multilayer deposits (work to appear).

**8. Concluding remarks**

This work has shown that non-Gaussian removal forces associated with the highly intermittent sweeping and ejection events in the near wall region of a turbulent boundary layer have a noticeable enhancement of monolayer resuspension rates compared to Gaussian forces; this is especially true in the long term erosion regime where removal rates are very small. The natural frequency of removal has a marked influence only on initial resuspension rates. For long term resuspension rates and removal fractions it has little effect. As a general observation, the large spread of adhesive forces generally occurring in practice and in experiment, reduces the influence of both non-Gaussian removal forces and the ratio of the rms to mean removal forces, most noticeable in the values of the removal fraction e.g. a factor of 2 in the value of the rms /mean removal force produces only a relatively small change ~ 0.15 in a value of 0.5 of the removal fraction for a geometric spread in adhesion ~8 (as in Hall's experiment for a nominally smooth substrate).

Finally we recall that while the focus of this work has been on resuspension from sparse monolayer coverages of particles, our eventual aim is to incorporate the non-Gaussian model for the 'primary' resuspension rate constant into a hybrid model for multilayer resuspension. Unlike monolayer resuspension, the timescale for removal of particles from multilayers plays a dominant role. What we have presented here is a necessary preliminary to the study of multilayer resuspension which we present in a subsequent paper.

**Acknowledgement**

The authors are grateful to Faouzi Laadhari and Richard Perkins (Ecole Centrale de Lyon) for the DNS data and valuable discussion of this work. This work was funded by the IRSN.